\documentclass[manuscript=article,journal=jctcce]{achemso}
\setkeys{acs}{articletitle = true}
\setkeys{acs}{maxauthors = 0}
\usepackage[T1]{fontenc}
\usepackage[utf8]{inputenc}
\usepackage{graphicx}
\usepackage{amsmath}
\usepackage{color}
\usepackage{comment}
\usepackage{braket}
\usepackage{hyperref}
\usepackage{multirow}
\usepackage{subfig}
\usepackage{amsmath,mathtools}
\usepackage[version=3]{mhchem}
\usepackage{longtable}
\usepackage{booktabs}
\usepackage{siunitx}
\usepackage{tablefootnote}
\usepackage{dcolumn}
\usepackage[normalem]{ulem}
\usepackage{cancel}
\usepackage[para,online]{threeparttable}

\definecolor{ao}{rgb}{1.0, 0.13, 0.32}
\definecolor{brightmagenta}{rgb}{0.90, 0.15, 0.55}
\newcommand{\at}[1]{\textcolor{brightmagenta}{#1}}

\title{Correcting basis set incompleteness in wave function correlation energy by dressing electronic Hamiltonian with 
an effective short-range interaction}

\author{Michał Hapka}
\affiliation{University of Warsaw, Faculty of Chemistry, ul.\ L.\ Pasteura 1, 02-093 Warsaw, Poland}

\author{Aleksandra Tucholska}
\affiliation{Institute of Physics, Lodz University of Technology, \mbox{ul.\ Wolczanska 217/221, 93-005 Lodz, Poland}}

\author{Marcin Modrzejewski}
\affiliation{University of Warsaw, Faculty of Chemistry, ul.\ L.\ Pasteura 1, 02-093 Warsaw, Poland}

\author{Pavlo Golub}
\affiliation{J. Heyrovsk\'{y} Institute of Physical Chemistry, Academy of Sciences of the Czech \mbox{Republic, v.v.i.}, Dolej\v{s}kova 3, 18223 Prague 8, Czech Republic}

\author{Libor Veis}
\affiliation{J. Heyrovsk\'{y} Institute of Physical Chemistry, Academy of Sciences of the Czech \mbox{Republic, v.v.i.}, Dolej\v{s}kova 3, 18223 Prague 8, Czech Republic}

\author{Katarzyna Pernal}
\email{pernalk@gmail.com}
\affiliation{Institute of Physics, Lodz University of Technology, \mbox{ul.\ Wolczanska 217/221, 93-005 Lodz, Poland}}


\begin{tocentry}
    \includegraphics[width=\textwidth]{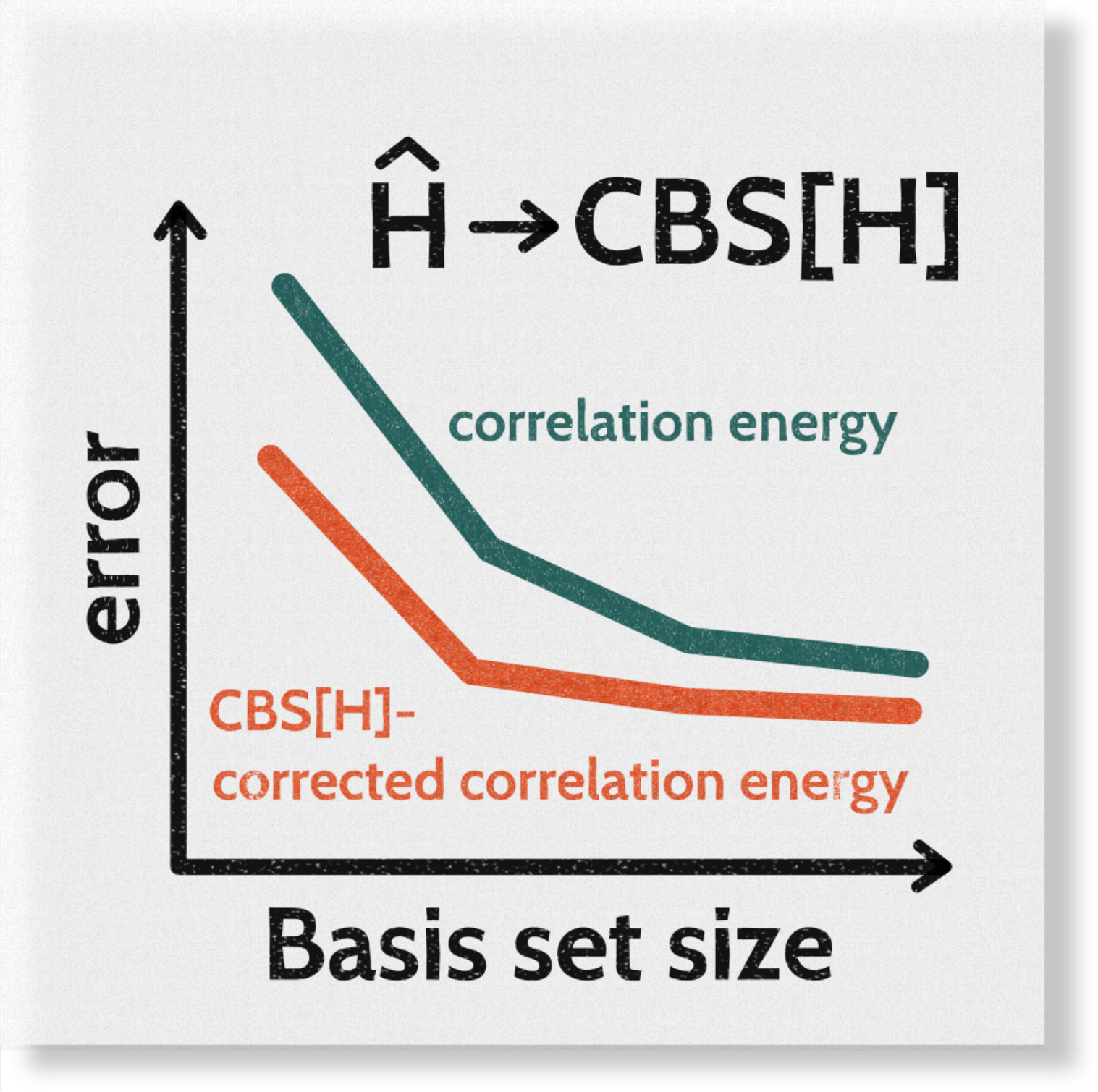}
  \end{tocentry}

\begin{document}


\begin{abstract}
 We propose a general approach to reducing basis set incompleteness error in electron correlation energy calculations. The correction is computed alongside the correlation energy in a single calculation by modifying the electron interaction operator with an effective short-range electron-electron interaction.
 Our approach is based on a local mapping between the Coulomb operator projected onto a finite basis and a long-range interaction represented by the error function with a local range-separated parameter, originally introduced by Giner et al. [J. Chem. Phys. 149, 194301 (2018)]. The complementary short-range interaction, included in the Hamiltonian, effectively accounts for the Coulomb interaction missing in a given basis.  As a numerical demonstration, we apply the method with complete active space wavefunctions. Correlation energies are computed using two distinct approaches: the  linearized adiabatic connection (AC0) method and n-electron valence state second-order perturbation theory (NEVPT2). We obtain encouraging results for the dissociation energies of test molecules, with accuracy in a triple-$\zeta$ basis set comparable to or exceeding that of uncorrected AC0 or NEVPT2 energies in a quintuple-$\zeta$ basis set. 
\end{abstract}  

\textit{Introduction.--}
Many-electron wavefunction theories (WFT) provide a powerful framework for predicting the physical and chemical properties of matter. However, their accuracy is inherently limited by the size of the one-electron basis set.
The steep computational cost of expanding the basis prevents reaching the so-called complete-basis-set (CBS) limit, which is essential for achieving quantitative agreement with experimental data. At the core of this problem is the slow convergence of the electron correlation energy with the basis set enlargement, which arises primarily from the difficulty in describing short-range correlation effects and the electron-electron cusp.

To address this issue, explicitly correlated (R12, F12) methods have been developed both for single-reference wavefunction approaches  \cite{werner2007general,fliegl2005coupled,werner2011explicitly} and multireference methods.~\cite{kong2012explicitly,shiozaki2013multireference} In the multireference context, these techniques have been incorporated into complete active space (CAS) multireference frameworks and their second-order perturbation corrections, leading to the development of CASPT2-F12\cite{shiozaki2010communication} and NEVPT2-F12\cite{guo2017explicitly} methodologies. Explicitly correlated methods significantly enhance the basis set convergence of the correlation energy, but their implementation comes with notable challenges. The need for an auxiliary basis set to evaluate three-electron integrals introduces additional computational overhead, and the underlying complex formalism requires substantial theoretical developments to integrate these methods into new computational frameworks.

An alternative approach to addressing the basis set incompleteness (BSI) error was proposed by Giner, Toulouse, and co-workers.~\cite{Giner:18} Their method, known as the density-based basis-set correction (DBBSC), is derived by matching the electron-electron Coulomb interaction projected onto a given basis set with a long-range electron interaction. The complementary short-range interaction defines the short-range correlation energy, which is equated to the BSI error. This error is then accounted for using approximate short-range correlation density functionals\cite{Toulouse:05,Ferte:19}. 
DBBSC has been applied to both single-reference\cite{Loos:19,hesselmann2024density} and multiconfigurational wavefunction methods\cite{Giner:19,Giner:20}. Although it does not outperform explicitly correlated methods, it offers reduced computational cost and memory requirements.\cite{mester2023basis,mester2024basis}

In this work, we introduce a novel approach to mitigate the BSI error in the electron correlation energy by modifying the Coulomb Hamiltonian. Specifically, we incorporate an effective, basis set-dependent short-range interaction designed to recover the missing short-range correlation effects arising from basis set limitations. Similar to DBBSC, the method employs an approximate mapping between the full Coulomb potential and a long-range interaction; however, unlike DBBSC, it does not rely on density functional approximations. We demonstrate the efficacy of the method by accelerating the convergence of dissociation energy computations using multireference wavefunction methods, showcasing its potential to improve electronic structure predictions for complex molecular systems.

\textit{Long-range electron interacting Hamiltonian and a real Hamiltonian in a finite basis set.--}
Let $\mathcal{B}$ indicate a finite (given) basis set of one-electron
functions. Assume that the orbitals contained in $\mathcal{B}$ are orthonormal.
The electronic Hamiltonian projected onto the Hilbert space spanned by basis set
functions in $\mathcal{B}$, in the second quantization is written as
\begin{equation}
\hat{H}_{\mathcal{B}} =\sum_{pq\in\mathcal{B}} \sum_{\sigma=\alpha,\beta
} h_{pq}\ \hat{a}_{p,\sigma}^{\dagger} \hat{a}_{q,\sigma} +
\frac{1}{2} \sum_{pqrs\in
\mathcal{B}}     \sum_{\sigma,\sigma^{\prime}=\alpha,\beta} g_{pqrs} \ \hat
{a}_{p,\sigma}^{\dagger} \hat{a}_{q,\sigma^{\prime}}^{\dagger} \hat{a}_{s,\sigma^{\prime}} \hat{a}_{r,\sigma} \ \ \ ,\label{HB}
\end{equation}
where $p,q,r,s$ correspond to spatial, real-valued, and orthonormal orbitals
$\left\{ \varphi_{p}(\mathbf{r})\right\}$; $\left\{h_{pq}\right\}$
denotes one-electron integrals involving kinetic energy ($t$) and nuclear-electron
($\upsilon_{\text{ne}}$) interaction,
$h_{pq} = t_{pq} + \left[ \upsilon_{\text{ne}}\right]_{pq}$,
and $\left\{g_{pqrs}\right\}$ are two-electron integrals for the Coulomb
electron interaction, namely
\begin{equation}
g_{pqrs} = \left\langle pq|\upsilon_{ee}|rs\right\rangle 
= \int\int \varphi_{p}(\mathbf{r}_{1}) \varphi_{q}(\mathbf{r}_{2}) r_{12}^{-1} \varphi_{r}(\mathbf{r}_{1}) \varphi_{s}(\mathbf{r}_{2}) \ \text{d}\mathbf{r}_{1} \text{d}\mathbf{r}_{2}\ \ \ . \label{gpqrs}
\end{equation}
The ground state energy corresponding to a finite-basis set wavefunction
$\Psi_{\mathcal{B}}$ obtained for the Hamiltonian $\hat{H}_{\mathcal{B}}$ differs from the nonrelativistic energy $\left\langle
\Psi|\hat{H}|\Psi\right\rangle $, where $\hat{H}$ is the full electronic Hamiltonian,
\begin{align}
\hat{H}  & =\hat{T}+\hat{V}_{\text{ne}}+\hat{V}_{ee}
\label{Hreal}
\end{align}
and $\Psi$ is the exact wavefunction. This
difference defines the BSI correction,
\begin{equation}
\varepsilon_{\mathcal{B}}^{\text{CBS}} = \left\langle \Psi|\hat{H}|\Psi\right\rangle -\left\langle \Psi_{\mathcal{B}}|\hat{H}_{\mathcal{B}}|\Psi_{\mathcal{B}}\right\rangle \ \ \ . \label{CBScorr}
\end{equation}
The correction is negative and its magnitude is related to a deficient description of the electron correlation at short interelectron distances. Thus, $\varepsilon_{\mathcal{B}}^{\text{CBS}}$ can be
regarded as a short-range correlation energy.

Consider a model Hamiltonian, indicated as $\hat{H}^{\text{LR}}$, which is free of singularity at electron coalescence, i.e.\ the Coulomb interaction, $\upsilon_{ee}\left(r_{12}\right)=r_{12}^{-1}$, is replaced by the long-range (LR) interaction operator
$\upsilon_{ee}^{\text{LR}}\left(  r_{12}\right)$. By definition, it is finite at $r_{12}=0$ and decays as $r_{12}^{-1}$ with inter-electron distance
\begin{align}
\lim_{r_{12}\rightarrow0}\ r_{12}\ \upsilon_{ee}^{\text{LR}}\left(
r_{12}\right)   &  =0\ \ \ ,\label{LR1}\\
\lim_{r_{12}\rightarrow \infty}\ r_{12}\ \upsilon_{ee}^{\text{LR}}\left( r_{12} \right) & 
=1 \ \ \ .\label{LR2}
\end{align}
The real-space representation and the second-quantized form of $\hat{H}^{\text{LR}}$ \at{,}
\begin{equation}
\hat{H}^{\text{LR}} = \sum_{pq}^{\text{CBS}} \sum_{\sigma=\alpha,\beta} \tilde
{h}_{pq}\ \hat{a}_{p,\sigma}^{\dagger} \hat{a}_{q,\sigma} + \frac{1}{2} \sum_{pqrs}^{\text{CBS}} \sum_{\sigma,\sigma^{\prime}=\alpha,\beta} g_{pqrs}^{\text{LR}}\ \hat{a}_{p,\sigma}^{\dagger} \hat{a}_{q,\sigma^{\prime}}^{\dagger} \hat{a}_{s,\sigma^{\prime}} \hat{a}_{r,\sigma}\ \ \ .\label{HLR}
\end{equation}
are  equivalent if a complete basis set  of one-electron
functions is assumed, which is indicated in the upper limits of summations in Eq.~\eqref{HLR}. Integrals $\{ g_{pqrs}^{\text{LR}} \}$ involve the long-range
electron interaction, namely 
$g_{pqrs}^{\text{LR}}=\left\langle pq|\upsilon_{ee}^{\text{LR}}|rs\right\rangle $.

In the one-electron part, the 
Hamiltonian $\hat{H}^{\text{LR}}$ includes
a local potential that fixes the ground state electron density 
$\rho_{\Psi^{\text{LR}}}(\mathbf{r})=\left\langle \Psi^{\text{LR}}|\hat{\rho
}(\mathbf{r})|\Psi^{\text{LR}}\right\rangle$ 
to be equal to the exact density $\rho
_{0}(\mathbf{r})$.
This potential is a  central concept in multi-configurational DFT \cite{Sav-INC-96,Stoll:1985,Toulouse:2004,Pollet:2002p2162,Pernal:22}. 
In one of such formulations\cite{gori2006properties,Toulouse:05,Ferte:19}, the energy is obtained as a sum of 
  the expectation value of the full Hamiltonian and the multideterminantal short-range (SR)  correlation functional\cite{Giner:18} \begin{equation}
E_{\text{c,md}}^{\text{SR}}[\rho] = \min_{\Psi\rightarrow\rho} \left\langle
\Psi|\hat{T}+\hat{V}_{ee}|\Psi\right\rangle 
- \left\langle \Psi^{\text{LR}}|\hat{T}+\hat{V}_{ee}|\Psi^{\text{LR}}\right\rangle \ \ \ .\label{srmd}
\end{equation}
This functional captures short-range correlation effects missing in
$\Psi^{\text{LR}}$.

Consider the wavefunctions $\Psi^{\text{LR}}$ and $\Psi_{\mathcal{B}}$ corresponding to the long-range Hamiltonian and the full Hamiltonian projected onto a finite basis, Eqs.~\eqref{HLR} and \eqref{HB}, respectively. Neither of these wavefunctions exhibits the electron coalescence cusp. Also, recall that by definition the operator $\upsilon_{ee}^{\text{LR}}\left(r_{12}\right)$ shares the long-range behavior of the Coulomb interaction, see Eq.(\ref{LR2}). It is therefore reasonable to
assume that if the basis set $\mathcal{B}$ is sufficiently large and the electron density obtained from $\Psi_{\mathcal{B}}$ is exact,
$\rho^{\mathcal{B}}(\mathbf{r})=\rho_{0}(\mathbf{r})$,
there exists a basis-set specific long-range interaction operator
$\upsilon_{ee}^{\text{LR,}\mathcal{B}}$, such that the  wavefunctions $\Psi^{\text{LR}}$ and $\Psi_{\mathcal{B}}$ yield the same energies with the real Hamiltonian
\begin{equation}
\left\langle \Psi^{\text{LR}}|\hat{T}+\hat{V}_{ee}|\Psi^{\text{LR}}\right\rangle =\left\langle \Psi_{\mathcal{B}}|\hat{T}+\hat{V}_{ee}|\Psi_{\mathcal{B}}\right\rangle \ \ \ . \label{CBSB}
\end{equation}
It has been used that electron densities are equal, and so are electron-nuclei interaction energies.
This assumption immediately leads to an equality between the basis-set incompleteness correction, Eq.~\eqref{CBScorr}, and the
multideterminantal SR correlation energy defined in Eq.~\eqref{srmd} and computed for $\rho^{\mathcal{B}}$,
\begin{equation}
\varepsilon_{\mathcal{B}}^{\text{CBS}} = E_{\text{c,md}}^{\text{SR}}
[\rho^{\mathcal{B}}]\ \ \ .\label{CBS12}
\end{equation}

The equivalence defined in Eq.~\eqref{CBS12} is at the heart of the Density-Based Basis-Set Correction\cite{Giner:18} which employs a particular construction of the long-range interaction $\upsilon_{ee}^{\text{LR}}$. In the first step, the electron-electron Coulomb operator projected onto the basis set $\mathcal{B}$,
see the second term in Eq.~\eqref{HB}, is  represented in real space by an effective interaction $W_{\Psi^{\mathcal{B}}}(\mathbf{r}_{1},\mathbf{r}_{2})$ given as
\begin{equation}
W_{\Psi^{\mathcal{B}}}(\mathbf{r}_1,\mathbf{r}_2) = 
\frac{\sum_{pqrstu} g_{pqtu}\ \Gamma_{pqrs}\ \varphi_{t}(\mathbf{r}_1) \varphi_u(\mathbf{r}_2) \varphi_r(\mathbf{r}_{1}) \varphi_s(\mathbf{r}_2)}{\rho^{(2)}(\mathbf{r}_1,\mathbf{r}_2)}\ \ \ ,
\label{WB}
\end{equation}
where the electron pair density function $\rho^{(2)}(\mathbf{r}_1,\mathbf{r}_2)$ is the diagonal part of the two-electron reduced density matrix $\Gamma$ obtained from the wavefunction $\Psi^{\mathcal{B}}$. In the second step, $W_{\Psi^{\mathcal{B}}}(\mathbf{r}_{1},\mathbf{r}_{2})$ is  mapped on the long-range electronic interaction 
given in terms of the error function
$\upsilon_{ee}^{\text{LR,}\mathcal{B}}(r_{12})=\operatorname{erf}
[\mu^{\mathcal{B}}r_{12}]r_{12}^{-1} $.
The mapping requires that locally $W_{\Psi^{\mathcal{B}}}(\mathbf{r}_{1},\mathbf{r}_{2})$ and $\upsilon_{ee}^{\text{LR,}\mathcal{B}}(r_{12})$ coincide at electron coalescence
which implies that a basis-set-specific range-separation parameter $\mu^{\mathcal{B}}$ acquires position dependence and takes the form
\begin{equation}
\mu^{\mathcal{B}}(\mathbf{r}) = \frac{\sqrt{\pi}}{2} W_{\Psi^{\mathcal{B}}}(\mathbf{r},\mathbf{r})\ \ \ .\label{mu}
\end{equation}
By construction, $\mu^{\mathcal{B}}(\mathbf{r})$ tends to infinity in the $\mathcal{B}\rightarrow $CBS limit
and the value of the basis set correction, Eq.~\eqref{CBS12}, vanishes.


\textit{Basis set incompleteness-corrected correlation energy.--}
Basis set-specific electron correlation energy is generally defined for a reference
wavefunction $\Psi_{\mathcal{B}}^{\text{ref}}$ as a difference between the exact energy in $\mathcal{B}$, given by the full configuration interaction (FCI) wavefunction $\Psi_{\mathcal{B}}^{\text{FCI}}$, and the reference energy
\begin{equation}
E_{\text{corr}}^{\mathcal{B}} =
\left\langle
\Psi_{\mathcal{B}}^{\text{FCI}}|\hat{H}_{\mathcal{B}}|\Psi_{\mathcal{B}
}^{\text{FCI}}\right\rangle -\left\langle \Psi_{\mathcal{B}}^{\text{ref}}
|\hat{H}_{\mathcal{B}}|\Psi_{\mathcal{B}}^{\text{ref}}\right\rangle
\ \ \ .\label{Ecorr}
\end{equation}
The BSI correction $\varepsilon_{\mathcal{B}}^{\text{CBS}}$, defined in Eq.~\eqref{CBScorr} with $\Psi=\Psi^{\text{FCI}}$ and $\Psi_{\mathcal{B}}=\Psi_{\mathcal{B}}^{\text{FCI}}$, recovers the exact energy 
$E_0= \left\langle \Psi^{\text{FCI}}|\hat{H}|\Psi^{\text{FCI}}\right\rangle$
\begin{equation}
E_0 = \left\langle \Psi_{\mathcal{B}}^{\text{ref}}|\hat{H}_{\mathcal{B}}
|\Psi_{\mathcal{B}}^{\text{ref}}\right\rangle +E_{\text{corr}}^{\mathcal{B}}
+\varepsilon_{\mathcal{B}}^{\text{CBS}}\ \ \ .
\end{equation}
Relying on the ideas put forward in Ref.\citenum{Giner:18}, below we propose a new basis set correction which is obtained together with the
electron correlation energy by a modification of the electron interaction operator. In our approach, a single computation recovers $E_{\text{corr}}^{\text{CBS},\mathcal{B}}$---the correlation energy corrected for the BSI error
\begin{equation}
E_{\text{corr}}^{\text{CBS},\mathcal{B}} = E_{\text{corr}}^{\mathcal{B}} + \varepsilon_{\mathcal{B}}^{\text{CBS}}\ \ \ .\label{EcorrCBS}
\end{equation}

As it has been discussed, the operators $\hat{H}^{\text{LR}}$ and $\hat
{H}_{\mathcal{B}}$ lead to short-range electron correlation reduced with respect to full Coulomb electron operator. Thus, it is justified to seek a basis-set specific long-range electron interaction operator, by definition satisfying the short and long-range limits given in Eqs.~\eqref{LR1} and \eqref{LR2}, which mimics the Coulomb operator projected onto a finite basis set $\mathcal{B}$. Our main
assumption is therefore that for a given basis set, there exists a long-range interaction operator $\upsilon_{ee}^{\text{LR},\mathcal{B}}$, whose action in the full Hilbert space is equivalent to that of the projected Coulomb operator\:
\begin{equation}
\sum_{pqrs\in\mathcal{B}}g_{pqrs}\ \hat{a}_{p}^{\dagger}\hat{a}_{q}^{\dagger
}\hat{a}_{s}\hat{a}_{r}=\sum_{pqrs}^{\text{CBS}}g_{pqrs}^{\text{LR,}%
\mathcal{B}}\ \hat{a}_{p}^{\dagger}\hat{a}_{q}^{\dagger}\hat{a}_{s}\hat{a}%
_{r}\ \ \ ,\label{CBSLR}%
\end{equation}
where the Coulomb integrals $g_{pqrs}$ are defined in Eq.~\eqref{gpqrs} and $g_{pqrs}^{\text{LR,}\mathcal{B}}=\left\langle pq|\upsilon_{ee}^{\text{LR},\mathcal{B}}|rs\right\rangle$.
The effective long-range\ interaction $\upsilon_{ee}^{\text{LR,}\mathcal{B}}\left(  r_{12}\right)$, determined by a given basis set $\mathcal{B}$, by
construction must be finite at electron coalescence and tend to the Coulomb interaction in the CBS limit:
\begin{equation}
\lim_{\mathcal{B}\rightarrow\text{CBS}}\ \ \ \upsilon_{ee}^{\text{LR}%
,\mathcal{B}}(r_{12})=\frac{1}{r_{12}}\ \ \ .
\end{equation}
Let us introduce a complementary basis set $\mathcal{\bar{B}}$, such that $\mathcal{B}\cup\mathcal{\bar{B}}=\text{CBS}$, and the SR interaction operator complementary to $\upsilon_{ee}^{\text{LR,}\mathcal{B}}$
\begin{align}
\upsilon_{ee}^{\text{SR,}\mathcal{B}}(r_{12}) &  =\frac{1}{r_{12}}%
-\upsilon_{ee}^{\text{LR,}\mathcal{B}}(r_{12})\ \ \ ,\\
g_{pqrs}^{\text{SR,}\mathcal{B}} &= g_{pqrs} - g_{pqrs}^{\text{LR,}\mathcal{B}}\ \ \ .
\label{glrsr}
\end{align}
From Eqs.~\eqref{LR1} and \eqref{LR2}, it follows that $\upsilon_{ee}^{\text{SR,}\mathcal{B}}(r_{12})$ is singular at $r_{12}=0$ and decays faster than $r_{12}^{-1}$ for large distances.
Employing in Eq.~\eqref{CBSLR} the straightforward relation
\begin{equation}
\sum_{pqrs\in\mathcal{B}}g_{pqrs}\ \hat{a}_{p}^{\dagger}\hat{a}_{q}^{\dagger
}\hat{a}_{s}\hat{a}_{r} 
= \sum_{pqrs}^{\text{CBS}}g_{pqrs}\ \hat{a}_{p}^{\dagger} \hat{a}_{q}^{\dagger} \hat{a}_{s} \hat{a}_{r} 
- \sum_{pqrs\in \mathcal{\bar{B}}} g_{pqrs}\ \hat{a}_{p}^{\dagger} \hat{a}_{q}^{\dagger} \hat
{a}_{s} \hat{a}_{r} \ \ \ ,
\end{equation}
where it should be understood that the operator on the left-hand side and the operator in the last term on the right-hand side act in orthogonal spaces,
together with Eq.~(\ref{glrsr})  leads to the equality of the Coulomb electron repulsion projected onto $\mathcal{\bar{B}}$ and the short-range interaction in the complete basis set limit
\begin{equation}
\sum_{pqrs\in\mathcal{\bar{B}}}g_{pqrs}\ \hat{a}_p^{\dagger} \hat{a}_q^\dagger \hat{a}_s \hat{a}_r 
= \sum_{pqrs}^{\text{CBS}} g_{pqrs}^{\text{SR,}\mathcal{B}}\ \hat{a}_p^\dagger \hat{a}_q^\dagger \hat{a}_s \hat{a}_r\ \ \ .\label{CBSSR}
\end{equation}
Thus, incompleteness of the Coulomb interaction projected onto a finite basis
set $\mathcal{B}$ is equivalent to a short-range interaction. 

Typically, a reference wavefunction $\Psi_{\mathcal{B}}^{\text{ref}}$, found
in a finite basis set $\mathcal{B}$, corresponds to a Hamiltonian $\hat
{H}_{\mathcal{B}}^{(0)}$, which includes one- and two-particle
operators.
The $\hat{H}_{\mathcal{B}}$ Hamiltonian, see Eq.~\eqref{HB}, used to obtain
the correlation energy $E_{\text{corr}}^{\mathcal{B}}$, see Eq.~\eqref{Ecorr}, is complemented by $\hat{H}_{\mathcal{B}}^{^{\prime}}$, i.e.
\begin{equation}
\hat{H}_{\mathcal{B}}^{^{\prime}}=\hat{H}_{\mathcal{B}}-\hat{H}_{\mathcal{B}}^{(0)} \ \ \ .\label{HpB}%
\end{equation}
Computation of the basis-set incompleteness error-free correlation energy
defined in Eq.~\eqref{EcorrCBS}, which can be written as $E_{\text{corr}}^{\text{CBS},\mathcal{B}}=\left\langle
\Psi^{\text{FCI}}|\hat{H}|\Psi^{\text{FCI}}\right\rangle -\left\langle
\Psi_{\mathcal{B}}^{\text{ref}}|\hat{H}_{\mathcal{B}}|\Psi_{\mathcal{B}%
}^{\text{ref}}\right\rangle $, is carried out with the exact Hamiltonian, Eq.~\eqref{Hreal}, of the following form
\begin{equation}
\hat{H} = \sum_{pq}^{\text{CBS}} \sum_{\sigma=\alpha,\beta}h_{pq}\ \hat
{a}_{p,\sigma}^{\dagger} \hat{a}_{q,\sigma} +\frac{1}{2} \sum_{pqrs}^{\text{CBS}} \sum_{\sigma,\sigma^{\prime}=\alpha,\beta} g_{pqrs}\ \hat{a}_{p,\sigma
}^{\dagger} \hat{a}_{q,\sigma^{\prime}}^\dagger \hat{a}_{s,\sigma^{\prime}} \hat{a}_{r,\sigma}\ \ \ .
\label{Hreal2}
\end{equation}
The difference between $\hat{H}$ and a model Hamiltonian $\hat{H}_{\mathcal{B}}^{(0)}$ reads
\begin{align}
\hat{H}^{\prime} &  =\hat{H} - \hat{H}_{\mathcal{B}}^{(0)} 
=\sum_{pq\in \mathcal{B}} h_{pq}^{\prime}\ \hat{a}_p^\dagger \hat{a}_q + \frac{1}{2}\sum_{pqrs\in\mathcal{B}} g_{pqrs}^{\prime}\ \hat{a}_p^\dagger \hat{a}_q^\dagger \hat{a}_s \hat{a}_r \nonumber \\
& +\sum_{pq\in\mathcal{\bar{B}}} h_{pq}\ \hat{a}_p^\dagger \hat{a}_q
+ \frac{1}{2}\sum_{pqrs\in\mathcal{\bar{B}}} g_{pqrs}\ \hat{a}_p^\dagger \hat{a}_q^\dagger \hat{a}_s \hat{a}_r
\end{align}
where $h_{pq}^\prime = h_{pq} - h_{pq}^{(0)}$ and $g_{pqrs}^\prime = g_{pqrs} - g_{pqrs}^{(0)} $.
Assuming that the BSI error primarily affects the description of the electron-electron cusp, while the one-electron functions (density and density matrix) are converged with respect to the basis set, allows us to neglect the  $\sum_{pq\in\mathcal{\bar{B}}}h_{pq}\ \hat{a}_{p}^{\dagger}\hat{a}_{q}$ operator. 
Using Eq.~\eqref{CBSSR}, we obtain
\begin{equation}
\hat{H}^{\prime}=\sum_{pq\in\mathcal{B}}h_{pq}^{\prime}\ \hat{a}_{p}^{\dagger
}\hat{a}_{q}+\frac{1}{2}\sum_{pqrs\in\mathcal{B}}g_{pqrs}^{\prime}\ \hat{a}_{p}^{\dagger
}\hat{a}_{q}^{\dagger}\hat{a}_{s}\hat{a}_{r}+\frac{1}{2}\sum_{pqrs}^{\text{CBS}}%
g_{pqrs}^{\text{SR,}\mathcal{B}}\ \hat{a}_{p}^{\dagger}\hat{a}_{q}^{\dagger
}\hat{a}_{s}\hat{a}_{r}\ \ \ .\label{HpSR0}%
\end{equation}
In the following, we assume that the basis set $\mathcal{B}$ is sufficiently large for most of the BSI correction to the correlation energy to be recovered by the short-range interaction $\upsilon_{ee}^{\text{SR,}\mathcal{B}}$ represented in $\mathcal{B}$. 
This leads to the final form of the dressed operator, defined within a given basis set
\begin{equation}
\hat{H}^{\prime}=\hat{H}_{\mathcal{B}}^{^{\prime}}+\frac{1}{2}\sum_{pqrs\in\mathcal{B}} g_{pqrs}^{\text{SR,}\mathcal{B}}\ \hat{a}_{p}^{\dagger} \hat{a}_{q}^{\dagger} \hat{a}_{s}\hat{a}_{r}\ \ \, \label{HpSR}
\end{equation}
which is a central achievement of this work. It shows that BSI correction in the correlation energy calculation can be achieved by modifying the interacting operator $\hat{H}_{\mathcal{B}}^{^{\prime}}$, used to compute
basis-set-specific correlation energy, through the addition of a 
short-range interaction. This modification effectively enhances short-range electron correlation.

\textit{Effective short-range correlation operator.--}
Recall that Giner and co-authors\cite{Giner:18} have represented the electron-electron Coulomb
operator projected onto the given basis set $\mathcal{B}$ by an effective
interaction $W_{\Psi^{\mathcal{B}}}(\mathbf{r}_1,\mathbf{r}_2)$, Eq.~\eqref{WB}, which has been approximated locally by a long-range interaction with the error function,
$W_{\Psi^{\mathcal{B}}}(\mathbf{r}_{1},\mathbf{r}_{2})\approx\frac
{\operatorname{erf}\left[  \mu_{\mathcal{B}}(r_{1})r_{12}\right]  }{r_{12}}$
[see Eqs.~(20), (27), and (39) in Ref.\citenum{Giner:18}]. The proposed mapping is approximate, but has been shown to perform well in the DBBSC scheme. Since the function $\frac{\operatorname{erf}\left[ \mu_{\mathcal{B}}(r_{1})r_{12}\right]}{r_{12}}$ accurately approximates the action of the Coulomb operator in a finite basis, it is expected to satisfy the assumption of Eq.~\eqref{CBSLR}. Using this form of the long-range operator leads to the following short-range interaction integrals
\begin{equation}
g_{pqrs}^{\text{SR,}\mathcal{B}}=\left\langle pq|\frac{1-\operatorname{erf}%
\left[  \mu_{\mathcal{B}}(\mathbf{r}_{1})r_{12}\right]  }{r_{12}%
}|rs\right\rangle \ \ \ ,
\end{equation}
with a local dependence of the range-separation parameter. 
However, correlation energy calculations involving integrals with a local parameter would involve a prohibitive computational cost. To avoid this problem, we propose to use a factorized-$\mu$ dependence function of the following form
\begin{equation}
\upsilon_{ee}^{\text{SR,}\mathcal{B}}(r_{12})=e^{-\mu^{\mathcal{B}}%
(\mathbf{r}_{1})}\frac{1-\operatorname{erf}(r_{12})}{r_{12}}%
\ \ \ . \label{veesr}%
\end{equation}
Such factorization preserves the key features of the original operator, $\frac {1-\operatorname{erf}\left[\mu(\mathbf{r}_1)r_{12}\right]}{r_{12}}$, namely the singularity at electron coalescence,
its more rapid decay with $r_{12}$ compared to the Coulomb operator,
and the exponential decay with $\mu^{\mathcal{B}}$.
$\upsilon_{ee}^{\text{SR,}\mathcal{B}}(r_{12})$ decays less rapidly with large $\mu^{\mathcal{B}}$ than $\frac{1-\operatorname{erf}\left[
\mu(\mathbf{r}_{1})r_{12}\right]  }{r_{12}}$, which decreases as $\exp
[-\mu^{\mathcal{B}}(\mathbf{r}_{1})^{2}r_{12}^{2}]$. However, numerical tests have shown that the performance of $\upsilon_{ee}^{\text{SR,}\mathcal{B}}(r_{12})$ deteriorates if the exponential function is replaced with $\exp[-\mu^{\mathcal{B}}(\mathbf{r}_{1})^{2}]$.

Using the factorized form of the short-range interaction introduced in Eq.~\eqref{veesr}, we exploit the approximate resolution of identity $\delta(\mathbf{r}_{1}-\mathbf{r}^{\prime
})\approx\sum_{t}\varphi_{t}(\mathbf{r}_{1})\varphi_{t}(\mathbf{r}^{\prime})$, and express the SR integrals as a product of the one- and two-electron matrices
\begin{equation}
g_{pqrs}^{\text{SR,}\mathcal{B}} = \left\langle pq|\upsilon_{ee}^{\text{SR,}
\mathcal{B}}(r_{12})|rs\right\rangle 
\approx \sum_{t}\left\langle p|e^{-\mu^{\mathcal{B}}(\mathbf{r)}}|t\right\rangle \left\langle
tq|\frac{1-\operatorname{erf}(r_{12})}{r_{12}}|rs\right\rangle
\ \ \ .\label{srinteg}
\end{equation}

In this work, we propose to construct the complementary Hamiltonian $\hat{H}'$, defined in Eq.~\eqref{HpSR}, using short-range integrals, given in Eq.~\eqref{srinteg}, computed with the local range-separation function defined by Eqs.~\eqref{WB} and \eqref{mu}. The integrals are symmetrized to impose the symmetries of the Coulomb integrals. 
Note that due to the locality of the $\mu^{\mathcal{B}}$ function, the proposed Hamiltonian $\hat{H}'$ leads to size-consistent BSI-corrected correlation energies.

The proposed framework is generally applicable to both single- and multireference correlation energy calculations. In this work, we apply it to remove the BSI errors of multireference correlation energies corresponding to complete active space (CAS) wavefunctions. The correlation energy is obtained by the linearized adiabatic connection (AC0) method\cite{ac_prl,Pastorczak:18a,Pastorczak:18b,pastorczak2019capturing,Beran:21,guo2024spinless} and the n-electron valence state second-order perturbation theory (NEVPT2)\cite{Angeli2001,angeli2001n,nevpt2,guo1,guo2}. In both methods, the $\hat{H}_{\mathcal{B}}^{(0)}$ operator is in the form of the Dyall's Hamiltonian\cite{Dyall1995,guo2024spinless}. To account for the BSI error, the perturbing Hamiltonian is modified by adding an effective short-range operator, as shown in Eq.~\eqref{HpSR}. The resulting AC0 and NEVPT2 correlation energies are denoted by AC0-CBS[H] and NEVPT2-CBS[H], respectively. 

\textit{Computational details.--}
Computation of the AC0 and AC0-CBS[H] energies requires CASSCF 1- and 2-electron reduced density matrices, which have been obtained from the Molpro\cite{Molpro:12} program. The calculations of the AC0 and AC0-CBS[H] correlation energies have been carried out using GammCor \cite{gammcor} program. For AC0-CBS[H], first the symmetrized short-range integrals, Eq.~\eqref{srinteg}, are computed with the local range-separated parameter $\mu^{\mathcal{B}}(\mathbf{r})$ constructed according to Eq.~\eqref{mu} and Eq.~\eqref{WB}. We have reduced the cost of computing the $\mu^{\mathcal{B}}(\mathbf{r})$ function by introducing Cholesky decomposition of Coulomb integrals in the calculation of the effective interaction at electron coalescence, $W_{\Psi^{\mathcal{B}}}(\mathbf{r},\mathbf{r})$, defined in Eq.~\eqref{WB} (for details, see the Appendix of Ref.\citenum{hapka2024self}). Matrix elements of the $\rm{exp}[-\mu^{\mathcal{B}}(\mathbf{r})]$ operator follow from the numerical integration. Molecular electron integrals were obtained with the \texttt{gammcor-integrals}\cite{gammcor-ints} library. NEVPT2 and NEVPT2-CBS[H] calculations were carried out with the PySCF code\cite{sun2020recent} with additional custom subroutines.

The AC0-CBS[H] and NEVPT2-CBS[H] results are compared with the AC0 and NEVPT2 energies corrected for the BSI error by adding the DBBSC correction given in Eq.~(\ref{CBS12}), computed using the PBE-based short-range functional from Ref.\citenum{Ferte:19}. The resulting energies are be denoted as AC0-DBBSC and NEVPT2-DBBSC. The DBBSC correction was implemented in GammCor\cite{gammcor}.

Valence active spaces were used in CASSCF computations for N$_2$, H$_2$O, and O$_2$ molecules: N$_2$: CAS(8,10), H$_2$O: CAS(8,6), O$_2$: CAS(8,6). For F$_2$, CAS(14,8) was adopted. Selected equilibrium bond lengths are: \ce{N2}: $ 2.070$ a.u., \ce{H2O}: $1.809$ a.u., \ce{O2}: $ 2.282$ a.u., \ce{F2}: $ 2.730$ a.u. Dissociation energies were computed for the following stretched-bond geometries: \ce{N2}: $10.000$ a.u., \ce{H2O}: $9.500$ a.u., \ce{O2}: $ 10.000$ a.u., and \ce{F2}$: 9.500$ a.u. Calculations were carried out using Dunning's cc-pV$X$Z basis sets.\cite{Dunning:89}

\textit{Results.--}
First, we investigate the accuracy of the BSI-corrected correlation energy $E_{\text{corr}}^{\text{CBS},\mathcal{B}}$ for the helium atom using Hartree-Fock (HF),
CAS(2,5)SCF and CAS(2,14)SCF reference wavefunctions. The correlation energy in a
given basis set, $E_{\text{corr}}^{\mathcal{B}}$, is computed according to Eq.~\eqref{Ecorr}. The $E_{\text{corr}}^{\text{CBS},\mathcal{B}}$ correlation energy, defined in Eq.~(\ref{EcorrCBS}), is found
following the approximate method proposed in this work. Thus, it consists in
taking a difference of the corrected FCI and the reference energies
$E_{\text{corr}}^{\text{CBS},\mathcal{B}}=
\left\langle \Psi_{\mathcal{B}}^{\text{FCI}}|\hat{H}[\mathcal{B},\Psi_{\mathcal{B}}^{\text{ref}}%
]|\Psi_{\mathcal{B}}^{\text{FCI}}\right\rangle -\left\langle \Psi_{\mathcal{B}}^{\text{ref}%
}|\hat{H}_{\mathcal{B}}|\Psi_{\mathcal{B}}^{\text{ref}}\right\rangle$
where $\Psi_{\mathcal{B}}^{\text{FCI}}$ is a full CI\ function that diagonalizes the following Hamiltonian [cf.\ Eq.~(\ref{HpB}) and (\ref{HpSR})]
\begin{align}
\hat{H}[\mathcal{B},\Psi_\mathcal{B}^{\text{ref}}]  & =\hat{H}_{\mathcal{B}}^{(0)}+\hat
{H}^{\prime}\nonumber\\
& =\hat{H}_{\mathcal{B}} + \frac{1}{2}\sum_{pqrs\in\mathcal{B}}g_{pqrs}^{\text{SR,}\mathcal{B}}\ \hat{a}_{p}^{\dagger} \hat{a}_{q}^{\dagger} \hat{a}_{s} \hat{a}_{r}\ \ \ .\label{HBPsi}
\end{align}
Clearly, the total energy, uncorrected for basis set incompleteness, computed as the sum of the reference energy $E_\mathcal{B}^{\text{ref}}=\left\langle \Psi
_{\mathcal{B}}^{\text{ref}}|\hat{H}_{\mathcal{B}}|\Psi_{\mathcal{B}%
}^{\text{ref}}\right\rangle$ and $E_{\text{corr}}^{\mathcal{B}}$, i.e.\ $E_\mathcal{B}^{\text{ref}}+E_{\text{corr}%
}^{\mathcal{B}}=\left\langle \Psi
_{\mathcal{B}}^{\text{FCI}}|\hat{H}_{\mathcal{B}}|\Psi_{\mathcal{B}%
}^{\text{FCI}}\right\rangle $, does not depend on the reference
function. Contrary to that, the energy corrected for BSI, $E_\mathcal{B}^{\text{ref}}+E_{\text{corr}}^{\text{CBS},\mathcal{B}}=\left\langle \Psi_{\mathcal{B}}^{\text{FCI}}|\hat{H}[\mathcal{B}%
,\Psi_{\mathcal{B}}^{\text{ref}}]|\Psi_{\mathcal{B}}^{\text{FCI}}\right\rangle $, will in
practice depend on both the basis set and the reference wavefunction. Notice that if the reference wavefunction is equal to the FCI\ function in a given basis set, then $E_{\text{corr}}^{\mathcal{B}}=0$, but the corrected correlation energy $E_{\text{corr}}^{\text{CBS},\mathcal{B}}$ will be different from zero.

\begin{figure}
\centering
 \begin{tabular}{c}
  \includegraphics[width=0.9\textwidth]{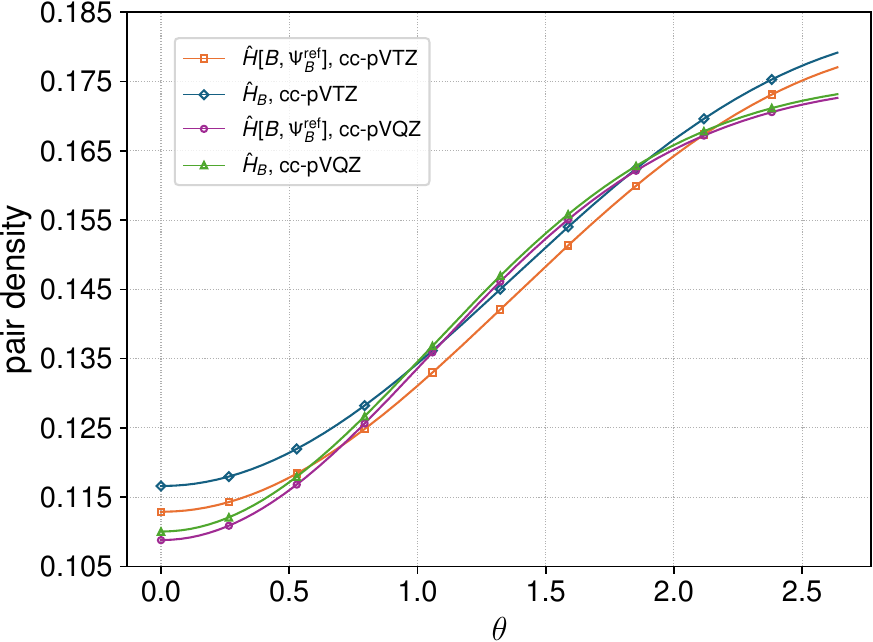}
 \end{tabular}
\caption{ 
Electron pair density on a sphere of radius 0.5 a.u.\ for the helium atom as a function of the angle between the position vectors of the two electrons. $\hat{H}_{\mathcal{B}}$ and 
$\hat{H}[\mathcal{B}, \Psi_\mathcal{B}^{\rm ref}]$ correspond to the unmodified Hamiltonian in a given basis set and the Hamiltonian modified according to Eq~(\ref{HBPsi}), respectively.
}
\label{fig_he}
\end{figure}

In Figure~\ref{fig_he}, we present electron pair densities obtained from the FCI wavefunction using both the unmodified Hamiltonian, $\hat{H}_{\mathcal{B}}$, and the Hamiltonian
$\hat{H}[\mathcal{B},\Psi_{\mathcal{B}}^{\text{ref}}]$, where $\Psi_{\mathcal{B}}^{\text{ref}}$ is a HF reference. For a given basis set, one observes
deepening of the pair function corresponding to $\hat{H}[\mathcal{B},\Psi_{\mathcal{B}}^{\text{ref}}]$ in the electron coalescence region, compared to  that obtained with the unmodified Hamiltonian. This confirms that the effective short-range operator deepens the Coulomb hole around the position of a reference electron. Its effect is therefore analogous to  increasing the basis set size---short-range electron correlation is strengthened.

In Table~\ref{tab_1} we report BSI corrections to the correlation energies, $\varepsilon_{\mathcal{B}}^{\text{CBS}}=E_{\text{corr}}%
^{\text{CBS},\mathcal{B}}-E_{\text{corr}%
}^{\mathcal{B}}$, Eq.~(\ref{EcorrCBS}),
for the helium atom obtained using the proposed approximate scheme, denoted as $\varepsilon
_{\mathcal{B}}^{\text{CBS[H]}}$. They are compared
with DBBSC results and the exact values.
The latter are computed following
Eq.~\eqref{CBScorr} as $E_{\text{exact}}-E_{\mathcal{B}}^{\text{FCI}}$, where
$E_{\mathcal{B}}^{\text{FCI}}=\left\langle \Psi_{\mathcal{B}}^{\text{FCI}%
}|\hat{H}_{\mathcal{B}}|\Psi_{\mathcal{B}}^{\text{FCI}}\right\rangle $ is the
FCI\ energy in a given basis set, and the exact energy for helium equal to
$-2.903724$ Ha is taken from Ref.\citenum{Davidson:91}.
The cc-pV$X$Z basis sets for $X$=(D, T, Q, 5, 6)\cite{Dunning:89} have been employed.

The poor performance of $\epsilon_{\mathcal{B}}^{\text{CBS[H]}}$  in the cc-pVDZ basis could be attributed to the
approximation applied in going from Eq.~\eqref{HpSR0} to Eq.~\eqref{HpSR}, i.e.\ projecting
the operator $\upsilon_{ee}^{\text{SR,}\mathcal{B}}$ onto the Hilbert space spanned by $\mathcal{B}$. The DBBSC correction is only slightly more accurate in this basis set. Expanding the basis set already to cc-pVTZ leads to a dramatic improvement in accuracy. The CBS[H] correction obtained for the CASSCF references agrees with the exact value to within 0.2~mHa. For larger basis sets, the error remains within the submillihartree regime  and reaches at most 0.5~mHa. A comparison with the DBBSC correction shows that both methods perform equally well on average. For a given basis set, both CBS[H] and DBBSC corrections are nearly independent on the reference CASSCF wavefunctions, as they should.  When applied with the HF reference, CBS[H] and DBBSC also perform on par, with the exception of the cc-pVTZ basis, where the CBS[H] error is 1~mHa larger than that of DBBSC. The HF-based errors are larger than for CASSCF wavefunctions, exceeding 0.5~mHa. 

\begin{table}
\centering
\begin{tabular}{c D{.}{.}{1} D{.}{.}{1} p{1ex} D{.}{.}{1}D{.}{.}{1} p{1ex} D{.}{.}{1}D{.}{.}{1} p{1ex} D{.}{.}{1}}
\hline
\multicolumn{1}{c}{\multirow{2}{*}{cc-pVXZ}} & \multicolumn{2}{c}{HF} && \multicolumn{2}{c}{CAS(2,5)SCF} && \multicolumn{2}{c}{CAS(2,14)SCF} && \multirow{2}{*}{exact} \\ & \multicolumn{1}{c}{$\epsilon_{\mathcal{B}}^{\text{CBS[H]}}$} &  \multicolumn{1}{c}{$\epsilon_{\mathcal{B}}^{\text{DBBSC}}$}  && \multicolumn{1}{c}{$\epsilon_{\mathcal{B}}^{\text{CBS[H]}}$} & \multicolumn{1}{c}{$\epsilon_{\mathcal{B}}^{\text{DBBSC}}$}  && \multicolumn{1}{c}{$\epsilon_{\mathcal{B}}^{\text{CBS[H]}}$} & \multicolumn{1}{c}{$\epsilon_{\mathcal{B}}^{\text{DBBSC}}$}  && \\ \hline
DZ & -9.0 & -12.3 && -8.3 & -9.5 && \multicolumn{1}{c}{-} & \multicolumn{1}{c}{-} && -16.1 \\
TZ & -3.9 &  -5.1 && -3.7 & -3.7 && -3.3 & -3.6 &&  -3.5 \\
QZ & -1.9 &  -2.3 && -1.8 & -1.6 && -1.5 & -1.8 &&  -1.3 \\
5Z & -1.1 &  -1.3 && -1.0 & -0.9 && -0.8 & -1.0 &&  -0.6 \\
6Z & -0.7 &  -0.8 && -0.6 & -0.5 && -0.6 & -0.5 &&  -0.3 \\ \hline
\end{tabular}
\caption{Basis set incompleteness corrections to correlation energies of the helium atom obtained using HF, CAS(2,5)SCF, and CAS(2,14)SCF reference wavefunctions. ``exact'' correction computed as $E_{\rm exact}-E^{\rm FCI}_{\mathcal{B}}$, where $E_{\rm exact}$ is taken from Ref.\citenum{Davidson:91}. Energy unit is mHa.}
\label{tab_1}
\end{table}

\begin{figure}
\centering
\includegraphics[width=\textwidth]{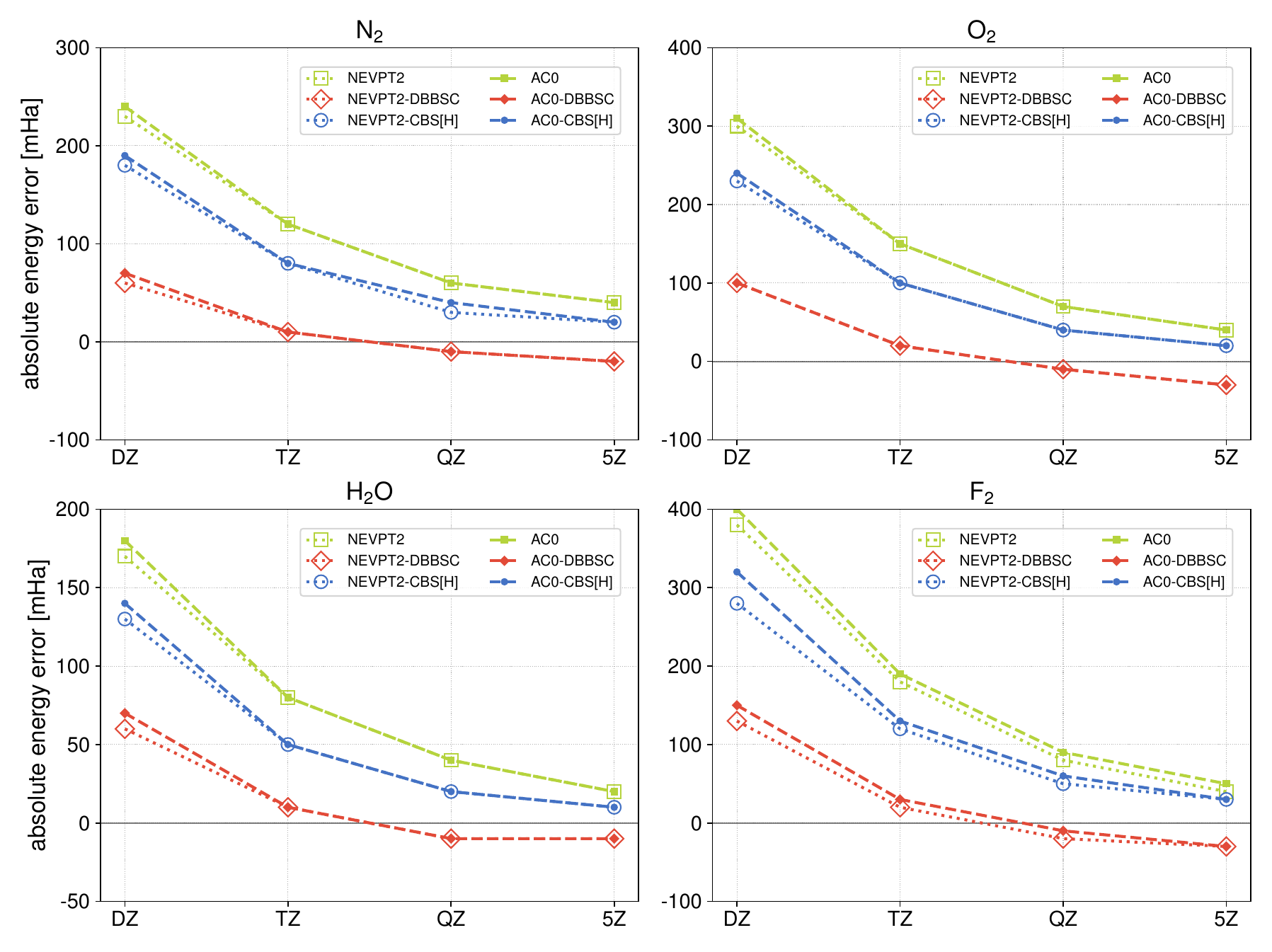}
\captionsetup{width=0.9\textwidth}
\caption{\footnotesize Absolute energy errors as a function of the cardinal number $X$ for N$_2$, O$_2$, H$_2$O and F$_2$ molecules in equilibrium geometries.
}
\label{fig2}
\end{figure}

The CBS[H] method was validated for absolute energies at equilibrium geometries and for relative dissociation energies on a set of representative molecules. Figure \ref{fig2} presents errors in the absolute energies computed with respect to the estimated benchmark values (see also Figure S1, Tables S1 and S2 in the Supporting Information). Benchmark results are calculated using a two-point extrapolation scheme\cite{halkier1998basis} from cc-pV5Z and cc-pV6Z values of the uncorrected correlation energy (AC0 or NEVPT2). In all cases,  the CBS[H]-corrected AC0 and NEVPT2 correlation energies computed in a basis set with cardinal number $X$ (starting from $X=3$) approach the uncorrected values obtained with ($X+1$) basis set.
Notably, the CBS[H] method is equally effective for AC0 and NEVPT2:  the improvement of convergence with the basis set size  is similar for both methods.  The DBBSC correction reduces the correlation energy error to a few mHa already in a triple-$\zeta$ basis. However, in QZ and 5Z basis sets, the DBBSC model overcorrects, and the energy error becomes negative. The CBS[H] does not exhibit this error.

\begin{figure}
\centering
\includegraphics[width=\textwidth]{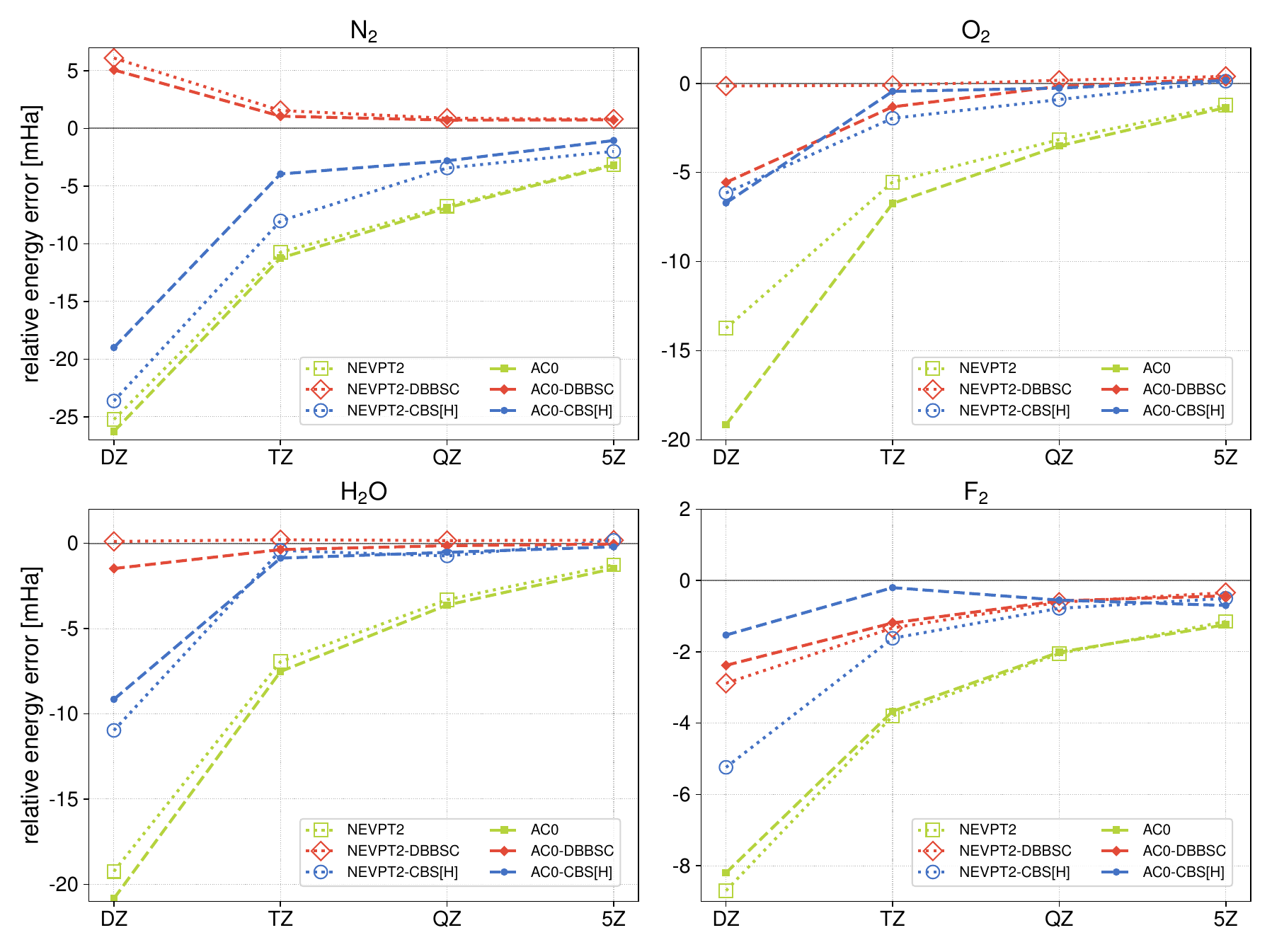}
\captionsetup{width=0.9\textwidth}
\caption{\footnotesize Errors of dissociation energies as a function of the cardinal number $X$ for N$_2$, O$_2$, H$_2$O and F$_2$ molecules.}
\label{fig3}
\end{figure}

In Figure~\ref{fig3} and Table \ref{tab2}, we present reduction of the BSI error in dissociation energies as a function of the cardinal number $X$. For the nitrogen molecule,  the AC0-CBS[H] energy in the TZ basis set recovers the uncorrected AC0 energy in the 5Z basis set to within 1~mHa. The NEVPT2-CBS[H] error in the TZ basis is larger and approaches that of the uncorrected NEVPT2 in the QZ basis. For the remaining molecules, the performance of AC0-CBS[H] is excellent and the dissociation energy error drops below 1~mHa already in the TZ basis set. Combined with NEVTP2, CBS[H] performs slightly worse, with errors for the O$_2$ and F$_2$ dimers exceeding 1~mHa at the TZ level. Still, the CBS[H] correction gains two cardinal numbers in accuracy relative to the uncorrected NEVPT2 energy. 
 On average, CBS[H] and DBBSC offer similar accuracy for relative correlation energies in basis sets larger than DZ. In the cc-pVTZ basis set, the DBBSC-corrected dissociation energies remain within an error margin of 1.5~mHa for all molecules. In CBS[H] calculations in the same basis, the nitrogen molecule is a clear outlier. On the other hand, for O$_2$ and F$_2$, AC0-CBS[H] achieves 0.5~mHa accuracy at the TZ level, whereas DBBSC errors exceed 1~mHa.

\begin{table}
\centering
\caption{Errors in dissociation energies as a function of the cardinal number $X$. Energy unit is mHa.}
 \begin{tabular}{cc SSS SSS} \hline
System & basis & \multicolumn{1}{c}{NEVPT2} & \multicolumn{1}{c}{NEVPT2} & \multicolumn{1}{c}{NEVPT2} & \multicolumn{1}{c}{AC0} & \multicolumn{1}{c}{AC0} & \multicolumn{1}{c}{AC0} \\
 &  & & \multicolumn{1}{c}{-DBBSC} & \multicolumn{1}{c}{-CBS[H]} &  & \multicolumn{1}{c}{-DBBSC} & \multicolumn{1}{c}{-CBS[H]} \\ \hline
\ce{N2}  & DZ & -25.23 & 6.10 & -23.63 & -26.27 & 5.07 & -19.00\\
 & TZ & -10.75 & 1.55 & -8.02 & -11.24 & 1.06 & -3.95 \\
 & QZ &  -6.73 & 0.88 & -3.43 &  -6.89 & 0.72 & -2.81 \\
 & 5Z &  -3.12 & 0.81 & -1.99 &  -3.19 & 0.74 & -1.05 \\ \hline
 \ce{H2O} & DZ & -19.24 & 0.16 & -10.97 & -20.83 & -1.48 & -9.15\\
 & TZ & -6.94 & 0.23 & -0.42 & -7.52 & -0.37 & -0.87\\
 & QZ & -3.31 & 0.19 & -0.74 & -3.62 & -0.14 & -0.53\\
 & 5Z & -1.26 & 0.19 & 0.17 & -1.48 & -0.04 & -0.21\\
 \midrule  
\ce{O2}  & DZ & -13.74 & -0.15 & -6.15 & -19.16 & -5.56 & -6.71\\
 & TZ & -5.54 & -0.11 & -1.96 & -6.74 & -1.32 & -0.45\\
 & QZ & -3.16 & 0.17 & -0.91 & -3.51 & -0.17 & -0.27\\
 & 5Z & -1.23 & 0.39 & 0.12 & -1.36 & 0.26 & 0.16\\
 \midrule  
\ce{F2} & DZ & -8.70 & -2.88 & -5.24 & -8.20 & -2.38 & -1.53\\
 & TZ & -3.80 & -1.33 & -1.62 & -3.67 & -1.19 & -0.20\\
 & QZ & -2.05 & -0.61 & -0.78 & -2.01 & -0.57 & -0.55\\
 & 5Z & -1.15 & -0.34 & -0.50 & -1.24 & -0.43 & -0.70\\ \hline
 \end{tabular}
\label{tab2}
\end{table}

In the DZ basis set, CBS[H] errors exceed 5~mHa, except for F$_2$. While this is still an improvement over the uncorrected  results, DBBSC in the same basis is clearly more accurate, with AC0-DBBSC errors ranging from $-1.5$ to 5.6~mHa. The superior performance of DBBSC over CBS[H] stems from a different construction of these two methods. By design, DBBSC accounts for the BSI error in both the reference and the correlation energies, while CBS[H], in its current formulation, corrects only the correlation energy.  Moreover, in a basis set as small as DZ, the resolution of identity assumption in Eq.~(\ref{srinteg}) underlying CBS[H] may be violated. To investigate this further, we carried out computations in the cc-pVXZ-F12 basis set\cite{peterson2008systematically}, optimized for explicitly correlated F12 methods. Results collected in the Supporting Information show that the errors of relative CBS[H]/cc-pVDZ-F12 energies are, on average, smaller than in the standard cc-pVDZ basis. Notably, for NEVPT2-CBS[H], they approach the values of uncorrected NEVPT2 in cc-pVTZ-F12 basis. 

\textit{Conclusions.--}
We have proposed the CBS[H] method to eliminate the basis set incompleteness error in the correlation energy. It is based on the assumption that a local mapping exists between the Coulomb electron interaction projected onto a finite basis set and a long-range interaction described by the error function with a local range-separation parameter. If this mapping holds, the corresponding short-range interaction is equivalent to the Coulomb interaction projected onto the space complementary to the chosen basis set. 
Building on the  construction of the long-range interaction with a local range-separation parameter  introduced in Ref.\citenum{Giner:18}, we have proposed an approximate form for the complementary interaction. The CBS[H] method  modifies the Hamiltonian used to compute the correlation energy  by adding an effective short-range interaction operator. As a result, the correlation energy corrected for BSI error is obtained in a single calculation.

Using the helium atom as a test case, we demonstrated that CBS[H] recovers the correlation energy with an accuracy better than 0.5~mHa in a triple-$\zeta$ basis set for both HF and CASSCF reference wavefunctions. We applied CBS[H] with the multireference AC0 and NEVPT2 models to compute dissociation energies for a set of representative small molecules. CBS[H] achieves an accuracy gain of approximately two cardinal numbers relative to the basis set size, starting from a triple-$\zeta$ basis. This is comparable to the performance of the DBBSC correction.\cite{Giner:18}

The computational cost of CBS[H] is comparable to on par with that of DBBSC, as both methods rely on evaluating a local range-separation parameter. With the Cholesky decomposition of two-electron integrals, the CBS[H] scaling is reduced to below the 5th power of the system size.  This ensures that the model does not introduce any additional overhead to the overall computational cost of correlation energy calculations.

A potential advantage of the method over DBBSC is its direct applicability to response properties, as the underlying dressing of the Hamiltonian with a short-range effective interaction is, in principle, universally valid. Future work will further explore broader applications of CBS[H].

\textit{Supporting Information.--}
Absolute energies in cc-pVXZ basis sets; absolute and relative energies in cc-pVXZ-F12 basis sets.

\textit{Acknowledgment.--}
This work has been supported by the National Science Center of Poland under grant no.\ 2021/43/I/ST4/02250 and the Czech Science Foundation (Grant No. 23-04302L).

\bibliography{articles}

\providecommand{\latin}[1]{#1}
\makeatletter
\providecommand{\doi}
  {\begingroup\let\do\@makeother\dospecials
  \catcode`\{=1 \catcode`\}=2 \doi@aux}
\providecommand{\doi@aux}[1]{\endgroup\texttt{#1}}
\makeatother
\providecommand*\mcitethebibliography{\thebibliography}
\csname @ifundefined\endcsname{endmcitethebibliography}
  {\let\endmcitethebibliography\endthebibliography}{}
\begin{mcitethebibliography}{44}
\providecommand*\natexlab[1]{#1}
\providecommand*\mciteSetBstSublistMode[1]{}
\providecommand*\mciteSetBstMaxWidthForm[2]{}
\providecommand*\mciteBstWouldAddEndPuncttrue
  {\def\EndOfBibitem{\unskip.}}
\providecommand*\mciteBstWouldAddEndPunctfalse
  {\let\EndOfBibitem\relax}
\providecommand*\mciteSetBstMidEndSepPunct[3]{}
\providecommand*\mciteSetBstSublistLabelBeginEnd[3]{}
\providecommand*\EndOfBibitem{}
\mciteSetBstSublistMode{f}
\mciteSetBstMaxWidthForm{subitem}{(\alph{mcitesubitemcount})}
\mciteSetBstSublistLabelBeginEnd
  {\mcitemaxwidthsubitemform\space}
  {\relax}
  {\relax}

\bibitem[Werner \latin{et~al.}(2007)Werner, Adler, and
  Manby]{werner2007general}
Werner,~H.-J.; Adler,~T.~B.; Manby,~F.~R. General orbital invariant MP2-F12
  theory. \emph{J. Chem. Phys.} \textbf{2007}, \emph{126}\relax
\mciteBstWouldAddEndPuncttrue
\mciteSetBstMidEndSepPunct{\mcitedefaultmidpunct}
{\mcitedefaultendpunct}{\mcitedefaultseppunct}\relax
\EndOfBibitem
\bibitem[Fliegl \latin{et~al.}(2005)Fliegl, Klopper, and
  H{\"a}ttig]{fliegl2005coupled}
Fliegl,~H.; Klopper,~W.; H{\"a}ttig,~C. Coupled-cluster theory with simplified
  linear-r12 corrections: The CCSD (R12) model. \emph{J. Chem. Phys.}
  \textbf{2005}, \emph{122}\relax
\mciteBstWouldAddEndPuncttrue
\mciteSetBstMidEndSepPunct{\mcitedefaultmidpunct}
{\mcitedefaultendpunct}{\mcitedefaultseppunct}\relax
\EndOfBibitem
\bibitem[Werner \latin{et~al.}(2011)Werner, Knizia, and
  Manby]{werner2011explicitly}
Werner,~H.-J.; Knizia,~G.; Manby,~F.~R. Explicitly correlated coupled cluster
  methods with pair-specific geminals. \emph{Mol. Phys.} \textbf{2011},
  \emph{109}, 407--417\relax
\mciteBstWouldAddEndPuncttrue
\mciteSetBstMidEndSepPunct{\mcitedefaultmidpunct}
{\mcitedefaultendpunct}{\mcitedefaultseppunct}\relax
\EndOfBibitem
\bibitem[Kong \latin{et~al.}(2012)Kong, Bischoff, and
  Valeev]{kong2012explicitly}
Kong,~L.; Bischoff,~F.~A.; Valeev,~E.~F. Explicitly correlated R12/F12 methods
  for electronic structure. \emph{Chem. Rev.} \textbf{2012}, \emph{112},
  75--107\relax
\mciteBstWouldAddEndPuncttrue
\mciteSetBstMidEndSepPunct{\mcitedefaultmidpunct}
{\mcitedefaultendpunct}{\mcitedefaultseppunct}\relax
\EndOfBibitem
\bibitem[Shiozaki and Werner(2013)Shiozaki, and
  Werner]{shiozaki2013multireference}
Shiozaki,~T.; Werner,~H.-J. Multireference explicitly correlated F12 theories.
  \emph{Mol. Phys.} \textbf{2013}, \emph{111}, 607--630\relax
\mciteBstWouldAddEndPuncttrue
\mciteSetBstMidEndSepPunct{\mcitedefaultmidpunct}
{\mcitedefaultendpunct}{\mcitedefaultseppunct}\relax
\EndOfBibitem
\bibitem[Shiozaki and Werner(2010)Shiozaki, and
  Werner]{shiozaki2010communication}
Shiozaki,~T.; Werner,~H.-J. Communication: Second-order multireference
  perturbation theory with explicit correlation: CASPT2-F12. \emph{J. Chem.
  Phys.} \textbf{2010}, \emph{133}\relax
\mciteBstWouldAddEndPuncttrue
\mciteSetBstMidEndSepPunct{\mcitedefaultmidpunct}
{\mcitedefaultendpunct}{\mcitedefaultseppunct}\relax
\EndOfBibitem
\bibitem[Guo \latin{et~al.}(2017)Guo, Sivalingam, Valeev, and
  Neese]{guo2017explicitly}
Guo,~Y.; Sivalingam,~K.; Valeev,~E.~F.; Neese,~F. Explicitly correlated
  N-electron valence state perturbation theory (NEVPT2-F12). \emph{J. Chem.
  Phys.} \textbf{2017}, \emph{147}\relax
\mciteBstWouldAddEndPuncttrue
\mciteSetBstMidEndSepPunct{\mcitedefaultmidpunct}
{\mcitedefaultendpunct}{\mcitedefaultseppunct}\relax
\EndOfBibitem
\bibitem[Giner \latin{et~al.}(2018)Giner, Pradines, Ferté, Assaraf, Savin, and
  Toulouse]{Giner:18}
Giner,~E.; Pradines,~B.; Ferté,~A.; Assaraf,~R.; Savin,~A.; Toulouse,~J.
  Curing basis-set convergence of wave-function theory using density-functional
  theory: A systematically improvable approach. \emph{J. Chem. Phys.}
  \textbf{2018}, \emph{149}, 194301\relax
\mciteBstWouldAddEndPuncttrue
\mciteSetBstMidEndSepPunct{\mcitedefaultmidpunct}
{\mcitedefaultendpunct}{\mcitedefaultseppunct}\relax
\EndOfBibitem
\bibitem[Toulouse \latin{et~al.}(2005)Toulouse, Gori-Giorgi, and
  Savin]{Toulouse:05}
Toulouse,~J.; Gori-Giorgi,~P.; Savin,~A. A short-range correlation energy
  density functional with multi-determinantal reference. \emph{Theor. Chem.
  Acc.} \textbf{2005}, \emph{114}, 305--308\relax
\mciteBstWouldAddEndPuncttrue
\mciteSetBstMidEndSepPunct{\mcitedefaultmidpunct}
{\mcitedefaultendpunct}{\mcitedefaultseppunct}\relax
\EndOfBibitem
\bibitem[Ferté \latin{et~al.}(2019)Ferté, Giner, and Toulouse]{Ferte:19}
Ferté,~A.; Giner,~E.; Toulouse,~J. Range-separated multideterminant
  density-functional theory with a short-range correlation functional of the
  on-top pair density. \emph{J. Chem. Phys.} \textbf{2019}, \emph{150},
  084103\relax
\mciteBstWouldAddEndPuncttrue
\mciteSetBstMidEndSepPunct{\mcitedefaultmidpunct}
{\mcitedefaultendpunct}{\mcitedefaultseppunct}\relax
\EndOfBibitem
\bibitem[Loos \latin{et~al.}(2019)Loos, Pradines, Scemama, Toulouse, and
  Giner]{Loos:19}
Loos,~P.-F.; Pradines,~B.; Scemama,~A.; Toulouse,~J.; Giner,~E. A Density-Based
  Basis-Set Correction for Wave Function Theory. \emph{J. Phys. Chem. Lett.}
  \textbf{2019}, \emph{10}, 2931--2937\relax
\mciteBstWouldAddEndPuncttrue
\mciteSetBstMidEndSepPunct{\mcitedefaultmidpunct}
{\mcitedefaultendpunct}{\mcitedefaultseppunct}\relax
\EndOfBibitem
\bibitem[He{\ss}elmann \latin{et~al.}(2024)He{\ss}elmann, Giner, Reinhardt,
  Knowles, Werner, and Toulouse]{hesselmann2024density}
He{\ss}elmann,~A.; Giner,~E.; Reinhardt,~P.; Knowles,~P.~J.; Werner,~H.-J.;
  Toulouse,~J. A density-fitting implementation of the density-based basis-set
  correction method. \emph{J. Comp. Chem.} \textbf{2024}, \emph{45},
  1247--1253\relax
\mciteBstWouldAddEndPuncttrue
\mciteSetBstMidEndSepPunct{\mcitedefaultmidpunct}
{\mcitedefaultendpunct}{\mcitedefaultseppunct}\relax
\EndOfBibitem
\bibitem[Giner \latin{et~al.}(2019)Giner, Scemama, Toulouse, and
  Loos]{Giner:19}
Giner,~E.; Scemama,~A.; Toulouse,~J.; Loos,~P.-F. Chemically accurate
  excitation energies with small basis sets. \emph{J. Chem. Phys.}
  \textbf{2019}, \emph{151}, 144118\relax
\mciteBstWouldAddEndPuncttrue
\mciteSetBstMidEndSepPunct{\mcitedefaultmidpunct}
{\mcitedefaultendpunct}{\mcitedefaultseppunct}\relax
\EndOfBibitem
\bibitem[Giner \latin{et~al.}(2020)Giner, Scemama, Loos, and
  Toulouse]{Giner:20}
Giner,~E.; Scemama,~A.; Loos,~P.-F.; Toulouse,~J. A basis-set error correction
  based on density-functional theory for strongly correlated molecular systems.
  \emph{J. Chem. Phys.} \textbf{2020}, \emph{152}, 174104\relax
\mciteBstWouldAddEndPuncttrue
\mciteSetBstMidEndSepPunct{\mcitedefaultmidpunct}
{\mcitedefaultendpunct}{\mcitedefaultseppunct}\relax
\EndOfBibitem
\bibitem[Mester and K{\'a}llay(2023)Mester, and K{\'a}llay]{mester2023basis}
Mester,~D.; K{\'a}llay,~M. Basis set limit of CCSD (T) energies: explicit
  correlation versus density-based basis-set correction. \emph{J. Chem. Theory
  Comput.} \textbf{2023}, \emph{19}, 8210--8222\relax
\mciteBstWouldAddEndPuncttrue
\mciteSetBstMidEndSepPunct{\mcitedefaultmidpunct}
{\mcitedefaultendpunct}{\mcitedefaultseppunct}\relax
\EndOfBibitem
\bibitem[Mester \latin{et~al.}(2024)Mester, Nagy, and
  K{\'a}llay]{mester2024basis}
Mester,~D.; Nagy,~P.~R.; K{\'a}llay,~M. Basis-set limit CCSD (T) energies for
  large molecules with local natural orbitals and reduced-scaling basis-set
  corrections. \emph{J. Chem. Theory Comput.} \textbf{2024}, \emph{20},
  7453--7468\relax
\mciteBstWouldAddEndPuncttrue
\mciteSetBstMidEndSepPunct{\mcitedefaultmidpunct}
{\mcitedefaultendpunct}{\mcitedefaultseppunct}\relax
\EndOfBibitem
\bibitem[Savin(1996)]{Sav-INC-96}
Savin,~A. In \emph{Recent Developments of Modern Density Functional Theory};
  Seminario,~J.~M., Ed.; Elsevier: Amsterdam, 1996; pp 327--357\relax
\mciteBstWouldAddEndPuncttrue
\mciteSetBstMidEndSepPunct{\mcitedefaultmidpunct}
{\mcitedefaultendpunct}{\mcitedefaultseppunct}\relax
\EndOfBibitem
\bibitem[Stoll and Savin(1985)Stoll, and Savin]{Stoll:1985}
Stoll,~H.; Savin,~A. In \emph{Density Functional Methods in Physics};
  Dreizler,~R., da~Providencia,~J., Eds.; Plenum, New York, 1985; pp
  177--207\relax
\mciteBstWouldAddEndPuncttrue
\mciteSetBstMidEndSepPunct{\mcitedefaultmidpunct}
{\mcitedefaultendpunct}{\mcitedefaultseppunct}\relax
\EndOfBibitem
\bibitem[Toulouse \latin{et~al.}(2004)Toulouse, Colonna, and
  Savin]{Toulouse:2004}
Toulouse,~J.; Colonna,~F.; Savin,~A. Long-range--short-range separation of the
  electron-electron interaction in density-functional theory. \emph{Phys. Rev.
  A} \textbf{2004}, \emph{70}, 062505\relax
\mciteBstWouldAddEndPuncttrue
\mciteSetBstMidEndSepPunct{\mcitedefaultmidpunct}
{\mcitedefaultendpunct}{\mcitedefaultseppunct}\relax
\EndOfBibitem
\bibitem[Pollet \latin{et~al.}(2002)Pollet, Savin, Leininger, and
  Stoll]{Pollet:2002p2162}
Pollet,~R.; Savin,~A.; Leininger,~T.; Stoll,~H. Combining multideterminantal
  wave functions with density functionals to handle near-degeneracy in atoms
  and molecules. \emph{J. Chem. Phys.} \textbf{2002}, \emph{116}, 1250\relax
\mciteBstWouldAddEndPuncttrue
\mciteSetBstMidEndSepPunct{\mcitedefaultmidpunct}
{\mcitedefaultendpunct}{\mcitedefaultseppunct}\relax
\EndOfBibitem
\bibitem[Pernal and Hapka(2022)Pernal, and Hapka]{Pernal:22}
Pernal,~K.; Hapka,~M. Range-separated multiconfigurational density functional
  theory methods. \emph{Wiley Interdiscip. Rev.: Comput. Mol. Sci.}
  \textbf{2022}, \emph{12}, e1566\relax
\mciteBstWouldAddEndPuncttrue
\mciteSetBstMidEndSepPunct{\mcitedefaultmidpunct}
{\mcitedefaultendpunct}{\mcitedefaultseppunct}\relax
\EndOfBibitem
\bibitem[Gori-Giorgi and Savin(2006)Gori-Giorgi, and Savin]{gori2006properties}
Gori-Giorgi,~P.; Savin,~A. Properties of short-range and long-range correlation
  energy density functionals from electron-electron coalescence. \emph{Phys.
  Rev. A} \textbf{2006}, \emph{73}, 032506\relax
\mciteBstWouldAddEndPuncttrue
\mciteSetBstMidEndSepPunct{\mcitedefaultmidpunct}
{\mcitedefaultendpunct}{\mcitedefaultseppunct}\relax
\EndOfBibitem
\bibitem[Pernal(2018)]{ac_prl}
Pernal,~K. Electron Correlation from the Adiabatic Connection for
  Multireference Wave Functions. \emph{Phys. Rev. Lett.} \textbf{2018},
  \emph{120}, 013001\relax
\mciteBstWouldAddEndPuncttrue
\mciteSetBstMidEndSepPunct{\mcitedefaultmidpunct}
{\mcitedefaultendpunct}{\mcitedefaultseppunct}\relax
\EndOfBibitem
\bibitem[Pastorczak and Pernal(2018)Pastorczak, and Pernal]{Pastorczak:18a}
Pastorczak,~E.; Pernal,~K. Correlation Energy from the Adiabatic Connection
  Formalism for Complete Active Space Wave Functions. \emph{J. Chem. Theory
  Comput.} \textbf{2018}, \emph{14}, 3493--3503\relax
\mciteBstWouldAddEndPuncttrue
\mciteSetBstMidEndSepPunct{\mcitedefaultmidpunct}
{\mcitedefaultendpunct}{\mcitedefaultseppunct}\relax
\EndOfBibitem
\bibitem[Pastorczak and Pernal(2018)Pastorczak, and Pernal]{Pastorczak:18b}
Pastorczak,~E.; Pernal,~K. Electronic Excited States from the
  Adiabatic-Connection Formalism with Complete Active Space Wave Functions.
  \emph{J. Phys. Chem. Lett.} \textbf{2018}, \emph{9}, 5534--5538\relax
\mciteBstWouldAddEndPuncttrue
\mciteSetBstMidEndSepPunct{\mcitedefaultmidpunct}
{\mcitedefaultendpunct}{\mcitedefaultseppunct}\relax
\EndOfBibitem
\bibitem[Pastorczak \latin{et~al.}(2019)Pastorczak, Hapka, Veis, and
  Pernal]{pastorczak2019capturing}
Pastorczak,~E.; Hapka,~M.; Veis,~L.; Pernal,~K. Capturing the Dynamic
  Correlation for Arbitrary Spin-Symmetry {CASSCF} Reference with Adiabatic
  Connection Approaches: Insights into the Electronic Structure of the
  Tetramethyleneethane Diradical. \emph{J. Phys. Chem. Lett.} \textbf{2019},
  \emph{10}, 4668--4674\relax
\mciteBstWouldAddEndPuncttrue
\mciteSetBstMidEndSepPunct{\mcitedefaultmidpunct}
{\mcitedefaultendpunct}{\mcitedefaultseppunct}\relax
\EndOfBibitem
\bibitem[Beran \latin{et~al.}(2021)Beran, Matoušek, Hapka, Pernal, and
  Veis]{Beran:21}
Beran,~P.; Matoušek,~M.; Hapka,~M.; Pernal,~K.; Veis,~L. Density matrix
  renormalization group with dynamical correlation via adiabatic connection.
  \emph{J. Chem. Theory Comput.} \textbf{2021}, \emph{17}, 7575--7585\relax
\mciteBstWouldAddEndPuncttrue
\mciteSetBstMidEndSepPunct{\mcitedefaultmidpunct}
{\mcitedefaultendpunct}{\mcitedefaultseppunct}\relax
\EndOfBibitem
\bibitem[Guo and Pernal(2024)Guo, and Pernal]{guo2024spinless}
Guo,~Y.; Pernal,~K. Spinless formulation of linearized adiabatic connection
  approximation and its comparison with second order N-electron valence state
  perturbation theory. \emph{Faraday Discuss.} \textbf{2024}, \emph{254},
  332--358\relax
\mciteBstWouldAddEndPuncttrue
\mciteSetBstMidEndSepPunct{\mcitedefaultmidpunct}
{\mcitedefaultendpunct}{\mcitedefaultseppunct}\relax
\EndOfBibitem
\bibitem[Angeli \latin{et~al.}(2001)Angeli, Cimiraglia, Evangelisti, Leininger,
  and Malrieu]{Angeli2001}
Angeli,~C.; Cimiraglia,~R.; Evangelisti,~S.; Leininger,~T.; Malrieu,~J.-P.
  Introduction of $n$-electron valence states for multireference perturbation
  theory. \emph{J. Chem. Phys.} \textbf{2001}, \emph{114}, 10252--10264\relax
\mciteBstWouldAddEndPuncttrue
\mciteSetBstMidEndSepPunct{\mcitedefaultmidpunct}
{\mcitedefaultendpunct}{\mcitedefaultseppunct}\relax
\EndOfBibitem
\bibitem[Angeli \latin{et~al.}(2001)Angeli, Cimiraglia, and
  Malrieu]{angeli2001n}
Angeli,~C.; Cimiraglia,~R.; Malrieu,~J.-P. N-electron valence state
  perturbation theory: a fast implementation of the strongly contracted
  variant. \emph{Chem. Phys. Lett.} \textbf{2001}, \emph{350}, 297--305\relax
\mciteBstWouldAddEndPuncttrue
\mciteSetBstMidEndSepPunct{\mcitedefaultmidpunct}
{\mcitedefaultendpunct}{\mcitedefaultseppunct}\relax
\EndOfBibitem
\bibitem[Angeli \latin{et~al.}(2002)Angeli, Cimiraglia, and Malrieu]{nevpt2}
Angeli,~C.; Cimiraglia,~R.; Malrieu,~J.-P. $n$-electron valence state
  perturbation theory: A spinless formulation and an efficient implementation
  of the strongly contracted and of the partially contracted variants. \emph{J.
  Chem. Phys.} \textbf{2002}, \emph{117}, 9138\relax
\mciteBstWouldAddEndPuncttrue
\mciteSetBstMidEndSepPunct{\mcitedefaultmidpunct}
{\mcitedefaultendpunct}{\mcitedefaultseppunct}\relax
\EndOfBibitem
\bibitem[Guo \latin{et~al.}(2021)Guo, Sivalingam, and Neese]{guo1}
Guo,~Y.; Sivalingam,~K.; Neese,~F. Approximations of density matrices in
  N-electron valence state second-order perturbation theory (NEVPT2). I.
  Revisiting the NEVPT2 construction. \emph{J. Chem. Phys.} \textbf{2021},
  \emph{154}, 214111\relax
\mciteBstWouldAddEndPuncttrue
\mciteSetBstMidEndSepPunct{\mcitedefaultmidpunct}
{\mcitedefaultendpunct}{\mcitedefaultseppunct}\relax
\EndOfBibitem
\bibitem[Guo \latin{et~al.}(2021)Guo, Sivalingam, Kollmar, and Neese]{guo2}
Guo,~Y.; Sivalingam,~K.; Kollmar,~C.; Neese,~F. Approximations of density
  matrices in N-electron valence state second-order perturbation theory
  (NEVPT2). II. The full rank NEVPT2 (FR-NEVPT2) formulation. \emph{J. Chem.
  Phys.} \textbf{2021}, \emph{154}, 214113\relax
\mciteBstWouldAddEndPuncttrue
\mciteSetBstMidEndSepPunct{\mcitedefaultmidpunct}
{\mcitedefaultendpunct}{\mcitedefaultseppunct}\relax
\EndOfBibitem
\bibitem[Dyall(1995)]{Dyall1995}
Dyall,~K.~G. The choice of a zeroth-order {Hamiltonian} for second-order
  perturbation theory with a complete active space self-consistent-field
  reference function. \emph{J. Chem. Phys.} \textbf{1995}, \emph{102},
  4909--4918\relax
\mciteBstWouldAddEndPuncttrue
\mciteSetBstMidEndSepPunct{\mcitedefaultmidpunct}
{\mcitedefaultendpunct}{\mcitedefaultseppunct}\relax
\EndOfBibitem
\bibitem[Werner \latin{et~al.}(2012)Werner, Knowles, Knizia, Manby, and
  Sch{\"u}tz]{Molpro:12}
Werner,~H.-J.; Knowles,~P.~J.; Knizia,~G.; Manby,~F.~R.; Sch{\"u}tz,~M. Molpro:
  a general-purpose quantum chemistry program package. 2012\relax
\mciteBstWouldAddEndPuncttrue
\mciteSetBstMidEndSepPunct{\mcitedefaultmidpunct}
{\mcitedefaultendpunct}{\mcitedefaultseppunct}\relax
\EndOfBibitem
\bibitem[Pernal \latin{et~al.}()Pernal, Hapka, Przybytek, Modrzejewski,
  Sokół, and Tucholska]{gammcor}
Pernal,~K.; Hapka,~M.; Przybytek,~M.; Modrzejewski,~M.; Sokół,~A.;
  Tucholska,~A. GammCor code. \url{https://github.com/pernalk/GAMMCOR},
  (accessed Jan 23, 2025)\relax
\mciteBstWouldAddEndPuncttrue
\mciteSetBstMidEndSepPunct{\mcitedefaultmidpunct}
{\mcitedefaultendpunct}{\mcitedefaultseppunct}\relax
\EndOfBibitem
\bibitem[Hapka \latin{et~al.}(2024)Hapka, Pastorczak, and
  Pernal]{hapka2024self}
Hapka,~M.; Pastorczak,~E.; Pernal,~K. Self-Adapting Short-Range Correlation
  Functional for Complete Active Space-Based Approximations. \emph{J. Phys.
  Chem. A} \textbf{2024}, \emph{128}, 7013--7022\relax
\mciteBstWouldAddEndPuncttrue
\mciteSetBstMidEndSepPunct{\mcitedefaultmidpunct}
{\mcitedefaultendpunct}{\mcitedefaultseppunct}\relax
\EndOfBibitem
\bibitem[Modrzejewski()]{gammcor-ints}
Modrzejewski,~M. gammcor-integrals library.
  \url{https://github.com/modrzejewski/gammcor-integrals}, (accessed Mar 28,
  2025)\relax
\mciteBstWouldAddEndPuncttrue
\mciteSetBstMidEndSepPunct{\mcitedefaultmidpunct}
{\mcitedefaultendpunct}{\mcitedefaultseppunct}\relax
\EndOfBibitem
\bibitem[Sun \latin{et~al.}(2020)Sun, Zhang, Banerjee, Bao, Barbry, Blunt,
  Bogdanov, Booth, Chen, Cui, \latin{et~al.} others]{sun2020recent}
Sun,~Q.; Zhang,~X.; Banerjee,~S.; Bao,~P.; Barbry,~M.; Blunt,~N.~S.;
  Bogdanov,~N.~A.; Booth,~G.~H.; Chen,~J.; Cui,~Z.-H.; others Recent
  developments in the PySCF program package. \emph{J. Chem. Phys.}
  \textbf{2020}, \emph{153}\relax
\mciteBstWouldAddEndPuncttrue
\mciteSetBstMidEndSepPunct{\mcitedefaultmidpunct}
{\mcitedefaultendpunct}{\mcitedefaultseppunct}\relax
\EndOfBibitem
\bibitem[Dunning~Jr(1989)]{Dunning:89}
Dunning~Jr,~T.~H. Gaussian basis sets for use in correlated molecular
  calculations. I. The atoms boron through neon and hydrogen. \emph{J. Chem.
  Phys.} \textbf{1989}, \emph{90}, 1007--1023\relax
\mciteBstWouldAddEndPuncttrue
\mciteSetBstMidEndSepPunct{\mcitedefaultmidpunct}
{\mcitedefaultendpunct}{\mcitedefaultseppunct}\relax
\EndOfBibitem
\bibitem[Davidson \latin{et~al.}(1991)Davidson, Hagstrom, Chakravorty, Umar,
  and Fischer]{Davidson:91}
Davidson,~E.~R.; Hagstrom,~S.~A.; Chakravorty,~S.~J.; Umar,~V.~M.;
  Fischer,~C.~F. Ground-state correlation energies for two- to ten-electron
  atomic ions. \emph{Phys. Rev. A} \textbf{1991}, \emph{44}, 7071--7083\relax
\mciteBstWouldAddEndPuncttrue
\mciteSetBstMidEndSepPunct{\mcitedefaultmidpunct}
{\mcitedefaultendpunct}{\mcitedefaultseppunct}\relax
\EndOfBibitem
\bibitem[Halkier \latin{et~al.}(1998)Halkier, Helgaker, J{\o}rgensen, Klopper,
  Koch, Olsen, and Wilson]{halkier1998basis}
Halkier,~A.; Helgaker,~T.; J{\o}rgensen,~P.; Klopper,~W.; Koch,~H.; Olsen,~J.;
  Wilson,~A.~K. Basis-set convergence in correlated calculations on Ne, N$_2$,
  and H$_2$O. \emph{Chem. Phys. Lett.} \textbf{1998}, \emph{286},
  243--252\relax
\mciteBstWouldAddEndPuncttrue
\mciteSetBstMidEndSepPunct{\mcitedefaultmidpunct}
{\mcitedefaultendpunct}{\mcitedefaultseppunct}\relax
\EndOfBibitem
\bibitem[Peterson \latin{et~al.}(2008)Peterson, Adler, and
  Werner]{peterson2008systematically}
Peterson,~K.~A.; Adler,~T.~B.; Werner,~H.-J. Systematically convergent basis
  sets for explicitly correlated wavefunctions: The atoms H, He, B--Ne, and
  Al--Ar. \emph{J. Chem. Phys.} \textbf{2008}, \emph{128}\relax
\mciteBstWouldAddEndPuncttrue
\mciteSetBstMidEndSepPunct{\mcitedefaultmidpunct}
{\mcitedefaultendpunct}{\mcitedefaultseppunct}\relax
\EndOfBibitem
\end{mcitethebibliography}

\end{document}



\section{Results in cc-pVXZ basis sets}
\begin{figure}[H]
    \centering
\includegraphics[width=\textwidth]{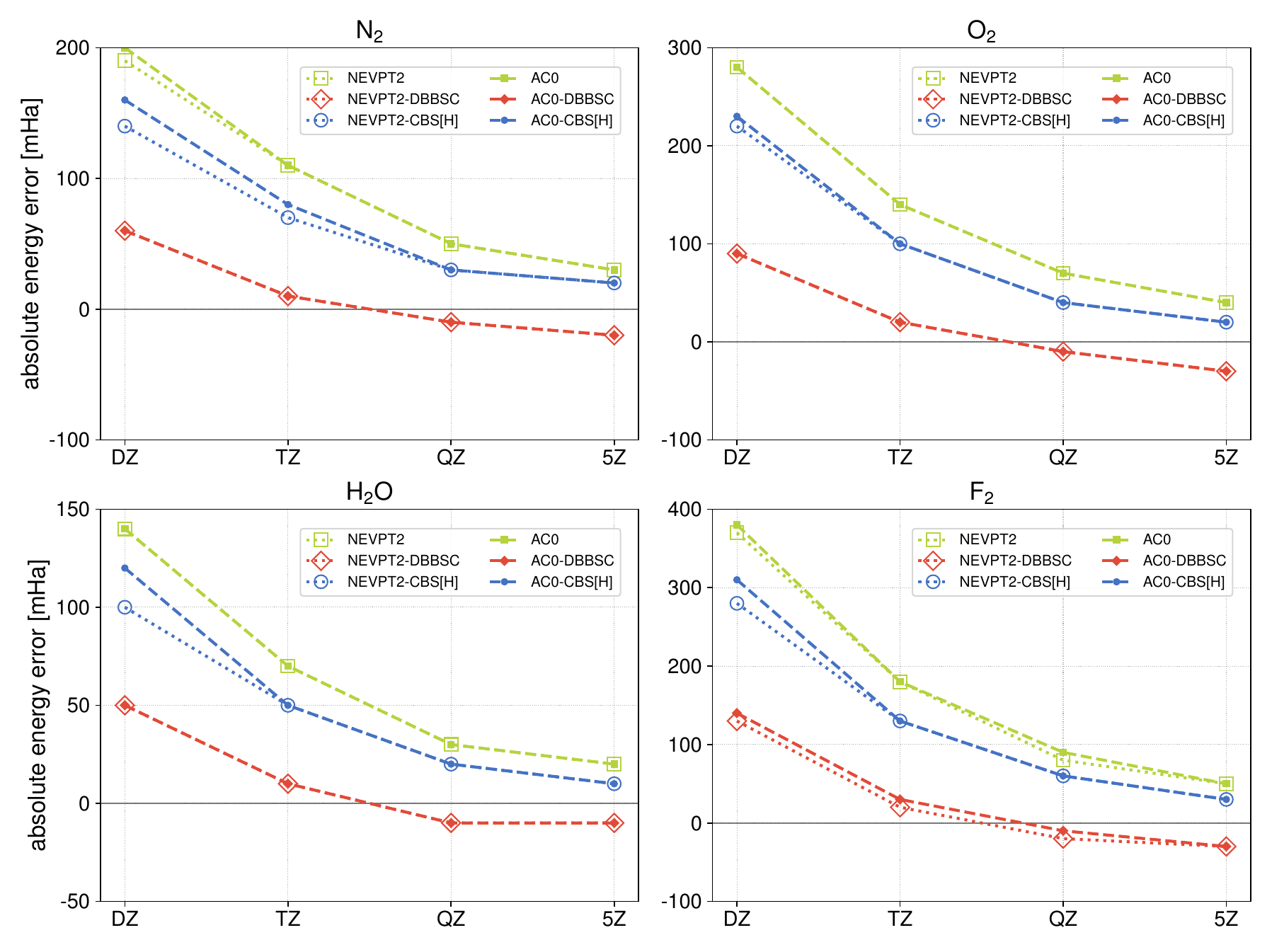}
                \captionsetup{width=0.9\textwidth}
        \caption{\footnotesize Absolute energy errors as a function of the cardinal number $X$ for molecules in dissociation geometries.}
  \figuretoc{Absolute energies in dissociation geometries} 
  \label{fig:absR}
\end{figure}
\begin{table}[H]
    \centering
    \renewcommand{\arraystretch}{1.4}
    \scriptsize
     \begin{threeparttable}
      \caption{\footnotesize Absolute energy errors for molecules in equilibrium geometries as a function of the cardinal number $X$. Energy unit is mHa. }
        \tabletoc{Absolute Energies  at $R_{eq}$} 
     \label{tab:absReq}
        \begin{tabular}{c cccc ccc }
    System&	basis&	NEVPT2	&NEVPT2-DBBSC&	NEVPT2-CBS[H]&	AC0&	AC0C-DBBSC	&AC0-CBS[H] \\
        \midrule
        \ce{N2} & DZ & 233.27 & 64.91 & 182.20 & 238.42 & 70.06 & 191.47\\
 & TZ & 122.15 & 12.34 & 83.67 & 123.73 & 13.93 & 84.41\\
 & QZ & 60.41 & -12.76 & 34.96 & 60.73 & -12.43 & 35.65\\
 & 5Z & 37.02 & -21.37 & 22.03 & 37.02 & -21.37 & 21.29\\
 \midrule
\ce{H2O} & DZ & 174.64 & 60.76 & 129.55 & 180.66 & 66.84 & 142.68\\
 & TZ & 81.46 & 11.14 & 51.65 & 83.47 & 13.17 & 55.00\\
 & QZ & 37.90 & -6.71 & 21.89 & 38.70 & -5.89 & 22.06\\
 & 5Z & 21.14 & -12.95 & 11.63 & 21.54 & -12.54 & 11.83\\
 \midrule
\ce{O2} & DZ & 304.42 & 102.48 & 232.48 & 306.51 & 104.57 & 244.44\\
 & TZ & 149.51 & 17.92 & 99.96 & 150.66 & 19.08 & 103.44\\
 & QZ & 71.00 & -14.38 & 40.89 & 71.53 & -13.86 & 42.49\\
 & 5Z & 41.09 & -25.79 & 22.78 & 41.34 & -25.53 & 23.69\\
 \midrule
\ce{F2} & DZ & 381.00 & 132.82 & 282.78 & 398.30 & 150.11 & 320.71\\
 & TZ & 177.89 & 19.34 & 124.58 & 186.58 & 28.04 & 133.99\\
 & QZ & 81.33 & -19.63 & 52.32 & 87.13 & -13.83 & 56.49\\
 & 5Z & 44.71 & -32.62 & 27.99 & 49.64 & -27.69 & 31.56\\
        \midrule   
    \end{tabular}
     \begin{tablenotes}
\scriptsize
        \item CAS used: 
        \ce{N2}-- CAS(8,10); 
        \ce{H2O} -- CAS(8,6);
        \ce{O2}--(8,6);
        \ce{F2}-- (14,8).
        \item $R_{eq}$ :
        \ce{N2}-- $ 2.07$ a.u.;
        \ce{H2O} -- $1.809$ a.u.;
         \ce{O2}--$ 2.282$ a.u.;
         \ce{F2}-- $ 2.730$ a.u.
    \end{tablenotes}
    \end{threeparttable}
\end{table}

\begin{table}[H]
    \centering
    \renewcommand{\arraystretch}{1.4}
    \scriptsize
     \begin{threeparttable}
      \caption{\footnotesize Absolute energy errors for molecules in dissociation geometries as a function of the cardinal number $X$. Energy unit is mHa.  }
        \tabletoc{Absolute Energies in dissociation geometries} 
     \label{tab:absR}
        \begin{tabular}{c cccc ccc }
    System&	basis&	NEVPT2	&NEVPT2-DBBSC&	NEVPT2-CBS[H]&	AC0&	AC0C-DBBSC	&AC0-CBS[H] \\
        \midrule
        \ce{N2} & DZ & 194.06 & 57.04 & 144.60 & 198.37 & 61.34 & 158.68\\
 & TZ & 108.64 & 11.13 & 72.89 & 109.91 & 12.40 & 77.89\\
 & QZ & 53.11 & -12.43 & 30.97 & 53.46 & -12.09 & 32.46\\
 & 5Z & 33.72 & -20.75 & 19.86 & 33.83 & -20.63 & 20.24\\
 \midrule
\ce{H2O} & DZ & 140.38 & 45.91 & 103.56 & 144.81 & 50.35 & 118.51\\
 & TZ & 71.02 & 7.86 & 47.72 & 72.45 & 9.30 & 50.63\\
 & QZ & 33.62 & -7.49 & 20.18 & 34.11 & -7.00 & 20.57\\
 & 5Z & 19.71 & -12.93 & 11.63 & 19.89 & -12.75 & 11.45\\
 \midrule
\ce{O2} & DZ & 281.81 & 91.83 & 217.02 & 278.75 & 88.77 & 228.82\\
 & TZ & 142.80 & 16.08 & 96.69 & 142.79 & 16.07 & 101.75\\
 & QZ & 67.65 & -14.77 & 39.94 & 67.83 & -14.59 & 42.08\\
 & 5Z & 39.74 & -25.66 & 22.93 & 39.85 & -25.55 & 23.83\\
 \midrule
\ce{F2} & DZ & 371.11 & 128.75 & 276.35 & 383.79 & 141.43 & 312.87\\
 & TZ & 179.06 & 22.99 & 127.94 & 182.77 & 26.70 & 133.65\\
 & QZ & 84.41 & -15.11 & 56.67 & 85.14 & -14.39 & 55.95\\
 & 5Z & 48.68 & -27.84 & 32.61 & 48.41 & -28.11 & 30.86\\
        \midrule   
    \end{tabular}
     \begin{tablenotes}
\scriptsize
        \item CAS used: 
        \ce{N2}-- CAS(8,10); 
        \ce{H2O} -- CAS(8,6);
        \ce{O2}--(8,6);
        \ce{F2}-- (14,8).
      \item $R_{x}$ :
         \ce{N2}-- $ 10.000$ a.u.;
        \ce{H2O} -- $9.500$ a.u.;
         \ce{O2}--$ 10.000$ a.u.;
         \ce{F2}-- $ 9.500$ a.u.
    \end{tablenotes}
    \end{threeparttable}
\end{table}

\section{Results in cc-pVXZ-F12 basis sets}
\begin{figure}[H]
    \centering
\includegraphics[width=\textwidth]{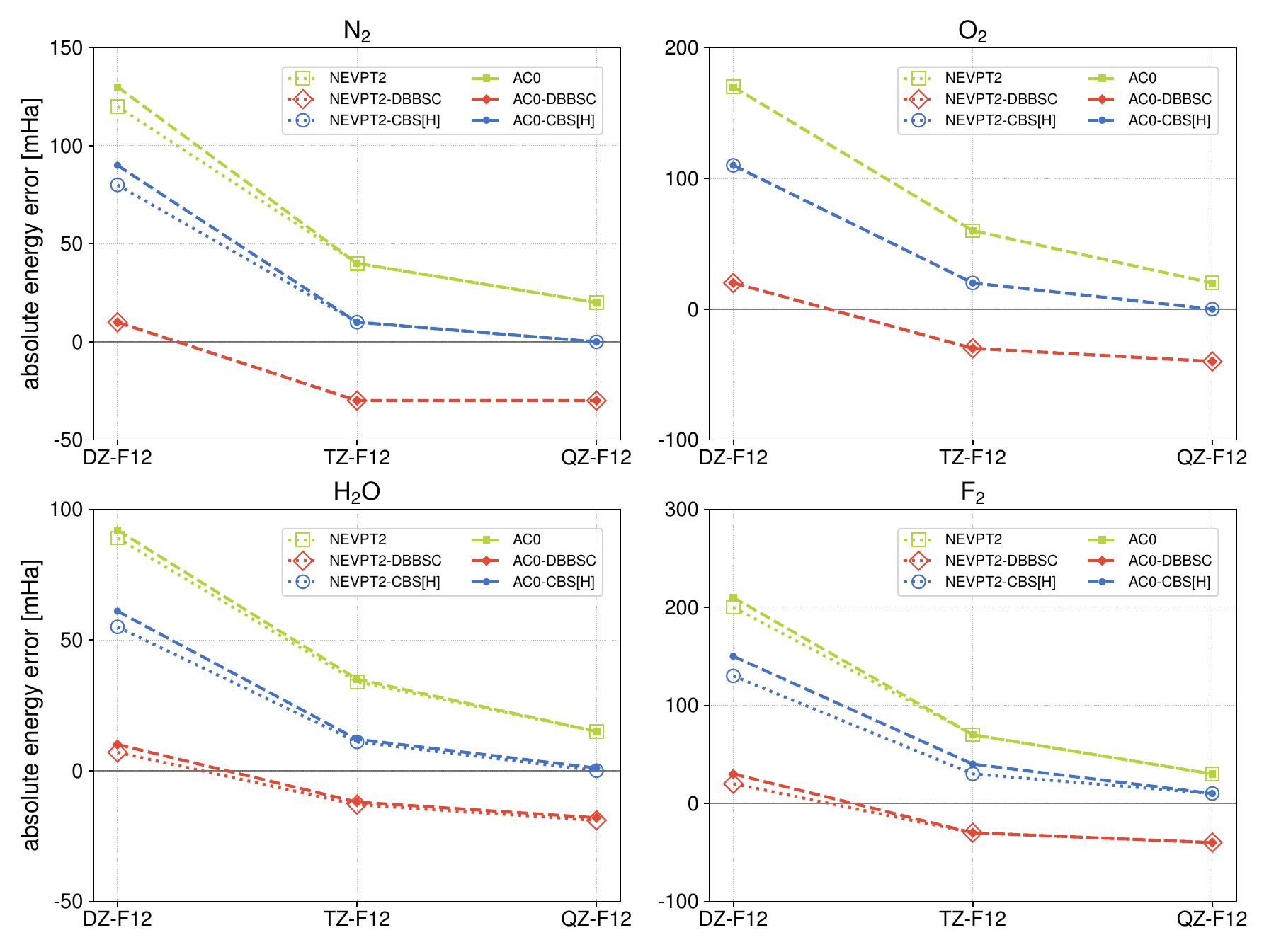}
                \captionsetup{width=0.9\textwidth}
        \caption{\footnotesize Absolute energy errors for molecules in equilibrium geometries as a function of the cardinal number $X$.}
        \figuretoc{Absolute energies at $R_{eq}$} 
  \label{fig:absReqf12}
\end{figure}
\begin{figure}[H]
    \centering
\includegraphics[width=\textwidth]{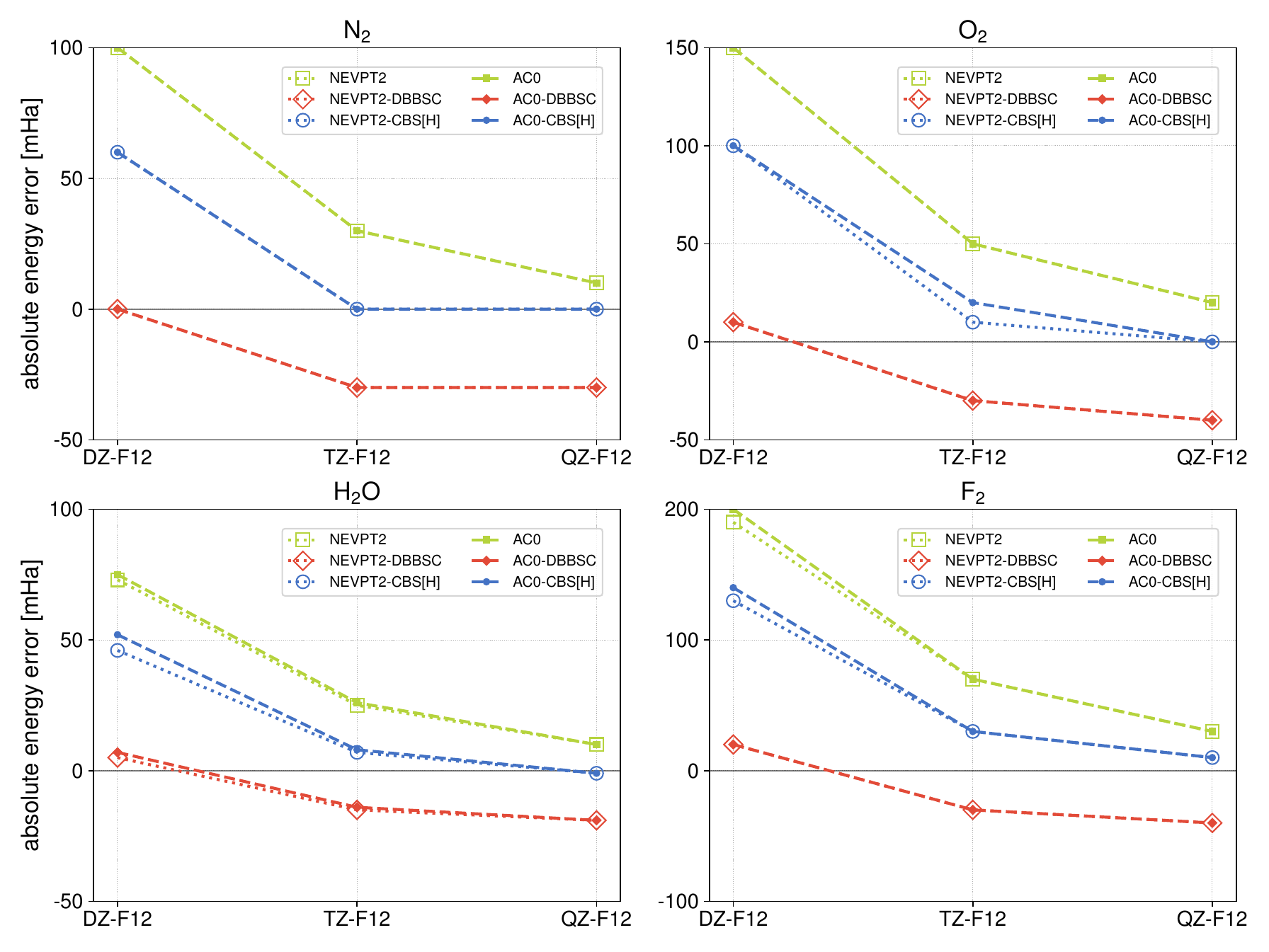}
                \captionsetup{width=0.9\textwidth}
        \caption{Absolute energy errors as a function of the cardinal number $X$ for molecules in dissociation geometries.}
        \figuretoc{Relative energies in dissociation geometries} 
  \label{fig:relRf12}
\end{figure}
\begin{figure}[H]
    \centering
\includegraphics[width=\textwidth]{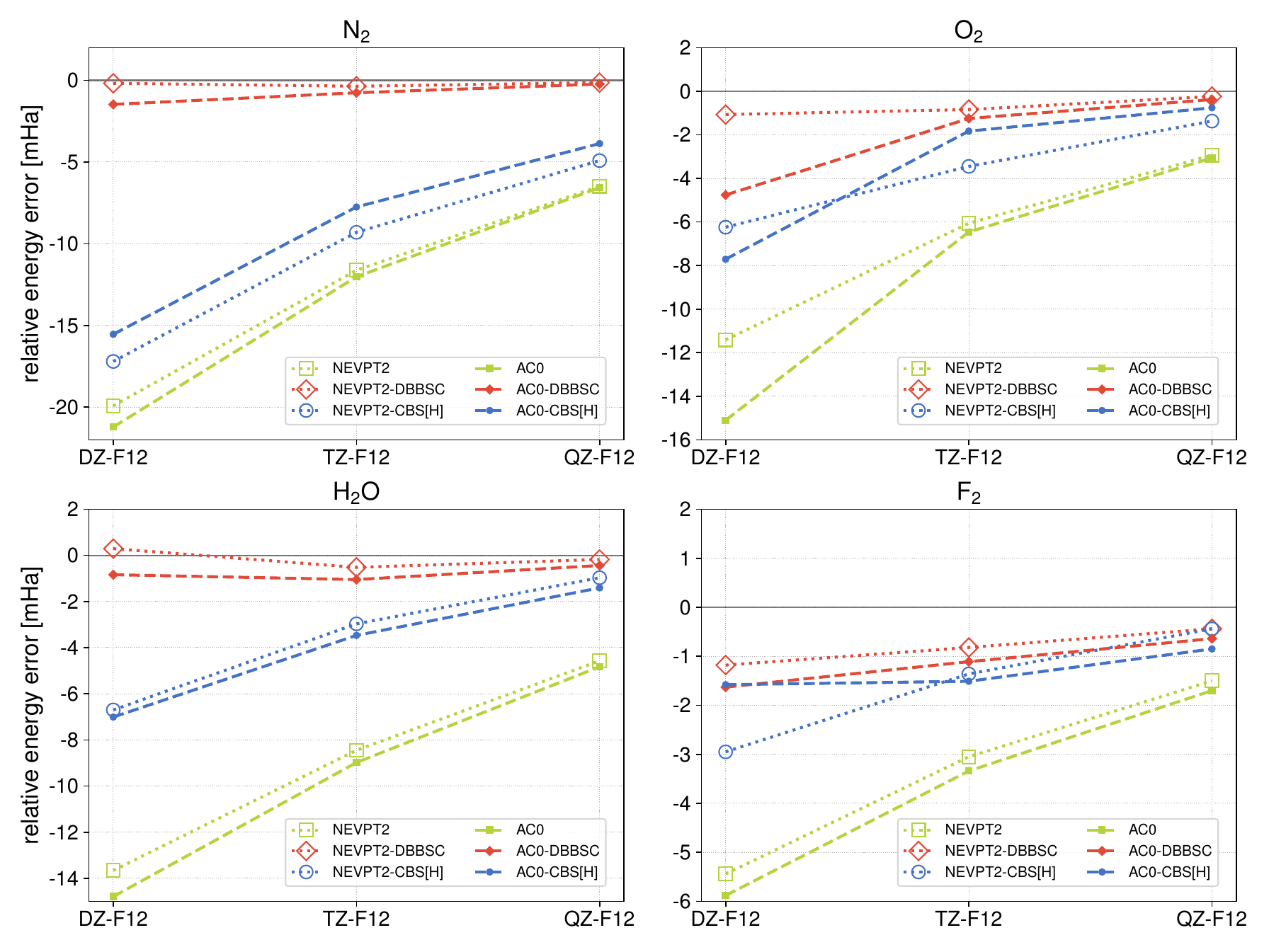}
                \captionsetup{width=0.9\textwidth}
        \caption{\footnotesize Errors of dissociation energies as a function of the cardinal number $X$.}
         \figuretoc{Relative energies at $R_{eq}$} 
  \label{fig:relReqf12}
\end{figure}

\begin{table}[H]
    \centering
    \renewcommand{\arraystretch}{1.4}
    \scriptsize
     \begin{threeparttable}
      \caption{\footnotesize Dissociation energy errors  as a function of the cardinal number $X$. Energy unit is mHa. }
      \tabletoc{Relative Energies at $R_{eq}$  } 
     \label{tab:relReqf12}
        \begin{tabular}{c cccc ccc }
    System&	basis&	NEVPT2	&NEVPT2-DBBSC&	NEVPT2-CBS[H]&	AC0&	AC0C-DBBSC	&AC0-CBS[H] \\
        \midrule
        \ce{N2} & DZ-F12 & -19.92 & -0.18 & -17.20 & -21.21 & -1.47 & -15.54\\
 & TZ-F12 & -11.62 & -0.36 & -9.30 & -12.02 & -0.76 & -7.75\\
 & QZ-F12 & -6.48 & -0.14 & -4.91 & -6.57 & -0.23 & -3.87\\
 \midrule
\ce{H2O} & DZ-F12 & -13.66 & 0.29 & -6.70 & -14.79 & -0.84 & -7.01\\
 & TZ-F12 & -8.45 & -0.52 & -2.97 & -8.98 & -1.05 & -3.47\\
 & QZ-F12 & -4.57 & -0.18 & -0.97 & -4.83 & -0.44 & -1.41\\
 \midrule
\ce{O2} & DZ-F12 & -11.41 & -1.07 & -6.24 & -15.10 & -4.76 & -7.71\\
 & TZ-F12 & -6.06 & -0.84 & -3.45 & -6.46 & -1.25 & -1.83\\
 & QZ-F12 & -2.93 & -0.24 & -1.37 & -3.06 & -0.38 & -0.76\\
 \midrule
\ce{F2} & DZ-f12 & -5.44 & -1.18 & -2.95 & -5.88 & -1.63 & -1.58\\
 & TZ-f12 & -3.05 & -0.82 & -1.36 & -3.34 & -1.11 & -1.51\\
 & QZ-f12 & -1.50 & -0.44 & -0.44 & -1.70 & -0.64 & -0.85\\
        \midrule   
    \end{tabular}
     \begin{tablenotes}
\scriptsize
        \item CAS used: 
        \ce{N2}-- CAS(8,10); 
        \ce{H2O} -- CAS(8,6);
        \ce{O2}--(8,6);
        \ce{F2}-- (14,8).
        \item $R_{eq}$ :
        \ce{N2}-- $ 2.07$ a.u.;
        \ce{H2O} -- $1.809$ a.u.;
         \ce{O2}--$ 2.282$ a.u.;
         \ce{F2}-- $ 2.730$ a.u.
         \item $R_{x}$ :
         \ce{N2}-- $ 10.000$ a.u.;
        \ce{H2O} -- $9.500$ a.u.;
         \ce{O2}--$ 10.000$ a.u.;
         \ce{F2}-- $ 9.500$ a.u.
    \end{tablenotes}
    \end{threeparttable}
    
\end{table}

\begin{table}[H]
    \centering
    \renewcommand{\arraystretch}{1.4}
    \scriptsize
     \begin{threeparttable}
      \caption{\footnotesize Absolute energy errors for molecules in equilibrium geometries as a function of the cardinal number $X$. Energy unit is mHa. }
         \tabletoc{Absolute Energies at $R_{eq}$  } 
     \label{tab:absReqf12}
        \begin{tabular}{c cccc ccc }
    System&	basis&	NEVPT2	&NEVPT2-DBBSC&	NEVPT2-CBS[H]&	AC0&	AC0C-DBBSC	&AC0-CBS[H] \\
        \midrule
        \ce{N2} & DZ-f12 & 121.72 & 5.71 & 80.80 & 125.06 & 9.05 & 85.21\\
 & TZ-f12 & 41.84 & -26.25 & 12.54 & 42.81 & -25.27 & 13.03\\
 & QZ-f12 & 19.42 & -31.30 & 0.41 & 19.79 & -30.93 & 0.11\\
 \midrule
\ce{H2O} & DZ-f12 & 89.23 & 6.99 & 55.29 & 92.34 & 10.10 & 60.73\\
 & TZ-f12 & 34.28 & -13.48 & 10.69 & 35.37 & -12.39 & 12.25\\
 & QZ-f12 & 14.62 & -18.64 & 0.26 & 15.07 & -18.19 & 0.58\\
 \midrule
\ce{O2} & DZ-f12 & 165.15 & 17.54 & 108.06 & 166.70 & 19.09 & 114.41\\
 & TZ-f12 & 57.60 & -27.65 & 18.42 & 58.24 & -27.01 & 20.45\\
 & QZ-f12 & 23.33 & -37.29 & -0.69 & 23.59 & -37.03 & 0.30\\
 \midrule
\ce{F2} & DZ-f12 & 201.09 & 19.02 & 133.63 & 207.44 & 25.37 & 146.35\\
 & TZ-f12 & 72.31 & -28.83 & 34.36 & 74.47 & -26.67 & 35.64\\
 & QZ-f12 & 32.25 & -40.99 & 8.81 & 33.20 & -40.04 & 8.57\\
        \midrule   
    \end{tabular}
     \begin{tablenotes}
\scriptsize
        \item CAS used: 
        \ce{N2}-- CAS(8,10); 
        \ce{H2O} -- CAS(8,6);
        \ce{O2}--(8,6);
        \ce{F2}-- (14,8).
        \item $R_{eq}$ :
        \ce{N2}-- $ 2.07$ a.u.;
        \ce{H2O} -- $1.809$ a.u.;
         \ce{O2}--$ 2.282$ a.u.;
         \ce{F2}-- $ 2.730$ a.u.
    \end{tablenotes}
    \end{threeparttable}
\end{table}

\begin{table}[H]
    \centering
    \renewcommand{\arraystretch}{1.4}
    \scriptsize
     \begin{threeparttable}
      \caption{\footnotesize Absolute energy errors for molecules in dissociation geometries as a function of the cardinal number $X$. Energy unit is mHa. }
        \tabletoc{Absolute Energies in dissociation geometries} 
     \label{tab:absRf12}
        \begin{tabular}{c cccc ccc }
    System&	basis&	NEVPT2	&NEVPT2-DBBSC&	NEVPT2-CBS[H]&	AC0&	AC0C-DBBSC	&AC0-CBS[H] \\
        \midrule
        \ce{N2} & DZ-f12 & 96.59 & 0.31 & 58.39 & 98.64 & 2.36 & 64.46\\
 & TZ-f12 & 29.79 & -27.04 & 2.81 & 30.37 & -26.46 & 4.85\\
 & QZ-f12 & 12.89 & -31.49 & -4.55 & 13.17 & -31.21 & -3.80\\
        \midrule
\ce{H2O} & DZ-f12 & 73.41 & 5.12 & 46.42 & 75.38 & 7.10 & 51.55\\
 & TZ-f12 & 25.13 & -14.69 & 7.02 & 25.69 & -14.14 & 8.08\\
 & QZ-f12 & 9.98 & -18.88 & -0.77 & 10.18 & -18.69 & -0.90\\
        \midrule
\ce{O2} & DZ-f12 & 148.73 & 11.45 & 96.81 & 146.59 & 9.32 & 101.69\\
 & TZ-f12 & 51.06 & -28.98 & 14.48 & 51.29 & -28.75 & 18.12\\
 & QZ-f12 & 20.40 & -37.54 & -2.07 & 20.53 & -37.41 & -0.46\\
        \midrule
\ce{F2} & DZ-f12 & 193.19 & 15.38 & 128.22 & 199.10 & 21.29 & 142.31\\
 & TZ-f12 & 69.04 & -29.88 & 32.78 & 70.91 & -28.01 & 33.90\\
 & QZ-f12 & 30.75 & -41.43 & 8.36 & 31.50 & -40.68 & 7.72\\
        \midrule   
    \end{tabular}
     \begin{tablenotes}
\scriptsize
        \item CAS used: 
        \ce{N2}-- CAS(8,10); 
        \ce{H2O} -- CAS(8,6);
        \ce{O2}--(8,6);
        \ce{F2}-- (14,8).
      \item $R_{x}$ :
         \ce{N2}-- $ 10.000$ a.u.;
        \ce{H2O} -- $9.500$ a.u.;
         \ce{O2}--$ 10.000$ a.u.;
         \ce{F2}-- $ 9.500$ a.u.
    \end{tablenotes}
    \end{threeparttable}
\end{table}